\documentclass[12pt]{iopart}

\usepackage[english]{babel}
\usepackage[dvips]{graphicx}
\usepackage{amsfonts}
\usepackage{amssymb}
\usepackage{mathrsfs}
\usepackage{epsfig}
\usepackage{subfigure}

\newcommand{\be}{\begin{eqnarray}}

\newcommand{\V}{\mathcal{V}}

\newcommand{\ee}{\end{eqnarray}}
\newcommand{\bdm}{\begin{displaymath}}
\newcommand{\edm}{\end{displaymath}}

\newcommand{\ba}{\begin{array}}
\newcommand{\ea}{\end{array}}
\newcommand{\Vds}{V_{\mathrm{dS}}}
\newcommand{\ps}{\widetilde{\psi}}
\newcommand{\Vk}{V_{\mathrm{KKLT}}}
\newcommand{\sd}{\sigma_{\mathrm{up}}}

\newcommand{\al}{\alpha'}

\newcommand{\ssi}{\sigma}
\newcommand{\hh}{h}

\newcommand{\co}{{\cal O}}

\newcommand{\kk}{\kappa_4}
\newcommand{\et}{\eta_{\,\mathbb{K}L\mathbb{M}T}}

\newcommand{\scr}{\sigma_{c}}
\newcommand{\tm}{t_{\mathrm{min}}}

\begin{document}

\title{Chasing Brane Inflation in String Theory}
\author{Axel Krause and Enrico Pajer}
\address{Arnold Sommerfeld Center for Theoretical Physics,
Department f\"ur Physik, Ludwig-Maximilians-Universit\"at M\"unchen,
Theresienstr.~37, 80333 M\"unchen, Germany}
\eads{\mailto{axel.krause@physik.uni-muenchen.de}, \mailto{enrico@theorie.physik.uni-muenchen.de}}

\begin{abstract}
We investigate the embedding of brane anti-brane inflation into a concrete type IIB string theory compactification with all moduli fixed. Specifically, we are considering a D3-brane, whose position represents the inflaton $\phi$, in a warped conifold throat in the presence of supersymmetrically embedded D7-branes and an anti D3-brane localized at the tip of the warped conifold cone. After presenting the moduli stabilization analysis for a general D7-brane embedding, we concentrate on two explicit models, the Ouyang and the Kuperstein embeddings. We analyze whether the forces, induced by moduli stabilization and acting on the D3-brane, might cancel by fine-tuning such as to leave us with the original Coulomb attraction of the anti D3-brane as the driving force for inflation. For a large class of D7-brane embeddings we obtain a negative result. Cancelations are possible only for very small intervals of $\phi$ around an inflection point but not globally. For the most part of its motion the inflaton then feels a steep, non slow-roll potential. We study the inflationary dynamics induced by this potential.
\end{abstract}
\noindent{\it Keywords:\/ string theory and cosmology, inflation\\
ArXiv ePrint: 0705.4682}

\maketitle


\section{Introduction}

One of the key steps to obtain a viable inflation scenario in string theory is to fix all massless moduli, except for the inflaton. In \cite{Giddings:2001yu}, in the framework of type IIB string compactifications, it was shown how fluxes can fix all complex structure moduli and the dilaton. The K\"ahler moduli enjoy a no-scale structure at tree level and therefore remain massless at this order. Quantum corrections, both perturbative and non-perturbative, however, break the no-scale structure.

In \cite{Kachru:2003aw} a three step procedure was proposed to fix, in addition, the K\"ahler moduli in positive energy vacua. First, one stabilizes the complex structure moduli and the dilaton by imposing the supersymmetry condition, $D_a W = 0$. Second, one considers non-perturbative effects, such as Euclidean D3-branes or gaugino condensation on a stack of D7-branes wrapping a divisor $\Sigma$ inside the Calabi-Yau threefold, which induce a K\"ahler moduli dependent term $W_{np}$ in the superpotential. This breaks the no-scale structure and produces supersymmetric anti-de Sitter (AdS) minima, obeying $D_iW=0$, where $i$ runs over the K\"ahler moduli. The final step is to uplift these AdS minima to non-supersymmetric de Sitter (dS) vacua in order to connect them to the real world.

Given that in principle all closed string moduli can thus be fixed in type IIB string compactification models, it is interesting to go one step further and introduce a suitable open string sector with the aim to model cosmic inflation using an open string modulus. One possibility is to identify the inflaton with the distance between a spacetime-filling mobile D3-brane and a fixed, very massive anti D3-brane (for recent reviews on this type of brane-antibrane inflation see \cite{Linde:2005dd}). The Coulomb attraction between the D3-brane and the anti D3-brane provides a potential which could drive inflation, provided the branes are located in a region with strong warping \cite{Kachru:2003sx}. Finally, one has to ensure that no other forces, in particular those which stabilize the K\"ahler moduli, spoil the achieved flatness of the potential. Unfortunately, this is generically the case \cite{Kachru:2003sx}. A non-trivial interplay between the volume and the D3-brane position moduli causes the K\"ahler moduli stabilization process to endow the inflaton with a mass of order the Hubble parameter, $H$. As a result, the second slow-roll parameter grows to $\eta \gtrsim 2/3$, showing the break-down of slow-roll inflation.

The idea of exactly canceling this moduli stabilization effect by some inflaton dependent threshold corrections to $W_{np}$, via fine-tuning, has received a certain amount of attention\footnote{It has been proposed in \cite{Pajer:2008uy} that also certain types of upliftings could be used to obtain this cancellation.}. Recently, threshold corrections to $W_{np}$ became available for the warped conifold background \cite{Baumann:2006th} (previously such effects had been calculated in \cite{Berg:2004ek} in the absence of warping). The result is that $W_{np}$ is proportional to the supersymmetric embedding $f(w)$ of the D7-branes to the power $1/n$. While $w$ collectively denotes the three complex coordinates of the D3-brane in the Calabi-Yau compactification space, $n$ represents the number of coinciding D7-branes in the stack on which gaugino condensation takes place ($n=1$ would apply to the Euclidean D3-brane case, which might alternatively be used for K\"ahler moduli stabilization). The zeros of $f(w)$ describe the embedding of the divisor $\Sigma$ which the D7-branes wrap.

It is one goal of this paper to analyze for concrete type IIB models whether all forces acting on the mobile D3-brane can add up to zero, except for the Coulomb attraction of the anti D3-brane which would then drive inflation. In the type IIB framework outlined above, we calculate the F-term potential for a general D7-brane embedding $f(w)$ and give general formulae for its moduli stabilization in the warped conifold background. Having stabilized all closed string moduli, we are left with an effective potential $V(r)$ for the radial position $r$ of the D3-brane in the warped conifold; this becomes the inflaton potential, $V(\phi)$, with the canonically normalized radial position of the D3-brane representing the inflaton. We then investigate inflation in the warped throat region by performing a small $\phi$ expansion of the potential.

The moduli stabilization effect that generically causes the break-down of slow-roll inflation, and that we would like to cancel by some additional mobile D3-brane dependence of $W_{np}$, induces a term proportional to $\phi^2$ in the potential. Unfortunately, no embedding allows for the creation of a compensating further term in the inflation potential with a $\phi^2$ dependence\footnote{Here we are assuming that inflation takes place far away from the tip of the conifold such that we can neglect the deformation parameter and use the singular conifold metric. It has been noticed in \cite{Pajer:2008uy} that using the exact deformed conifold metric, very close to the tip the moduli stabilization induces a term proportional to $\phi^3$ instead of $\phi^2$. This could in principle be canceled by the threshold corrections to the non-perturbative superpotential we are considering here. However, the cancellation would be valid only very close to the tip.\label{r3}}. In fact, the holomorphicity of the D7-brane embedding allows only integer powers of $\phi^{3/2}$ (some multiplied by a $\phi$ coming from the inverse conifold metric). This is crucial because terms with a different $\phi$ dependence can cancel only locally in a small $\phi$ interval, rather than globally. Outside this small interval the inflaton potential is not of the slow-roll type. Equivalently, outside this interval and despite fine-tuning, the motion of the D3-brane is governed by moduli stabilization effects compared to which the Coulomb potential represents only a subleading correction.

\begin{figure}
\includegraphics[height=6cm,width=0.45\textwidth]{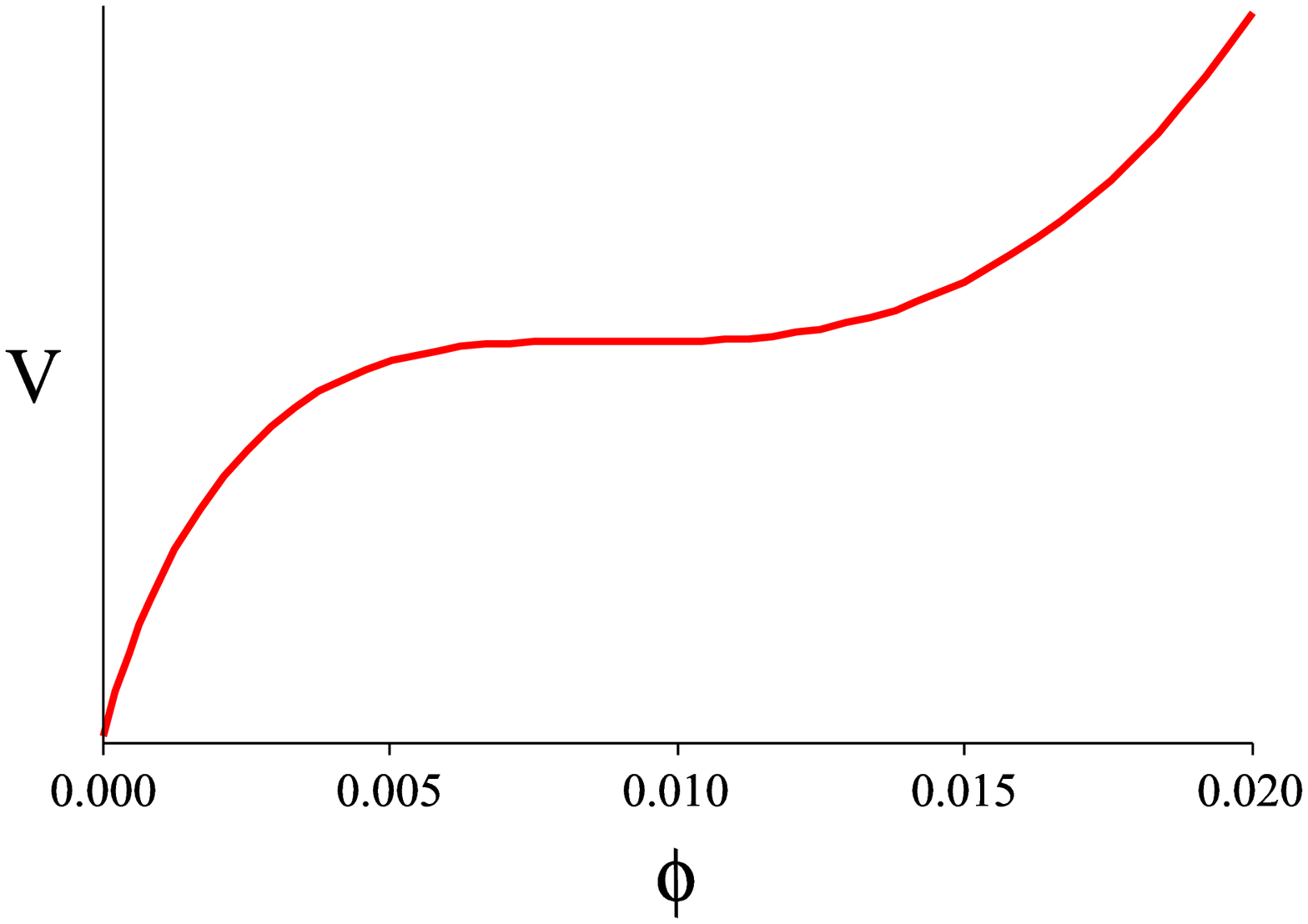}
\includegraphics[height=6cm,width=0.45\textwidth]{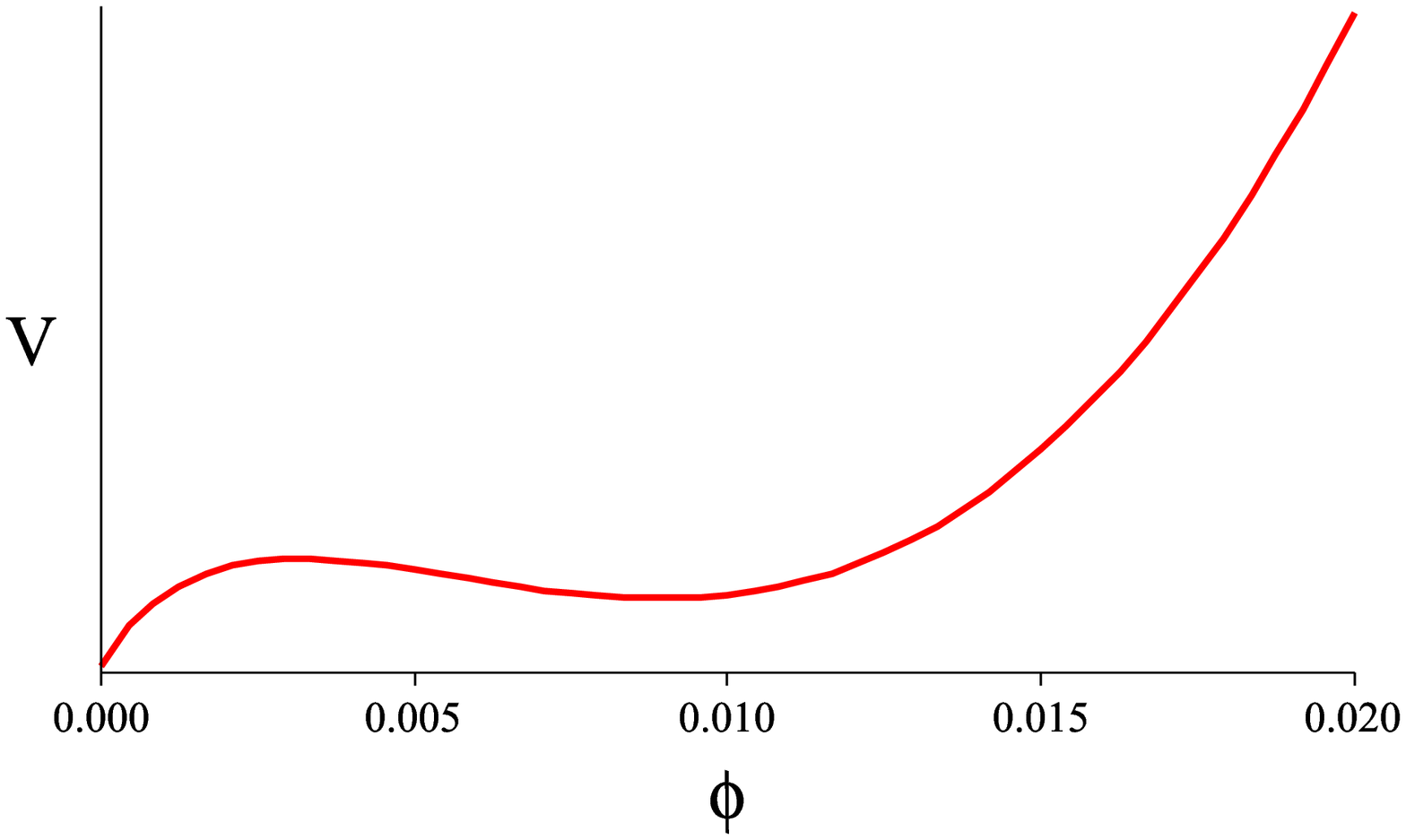}
\caption{\label{twins}The plots display the inflaton potential $V(\phi)$ for the Kuperstein embedding for two different values of the uplifting parameter $\beta=1.21$ (left) $\beta=1.4$ (right). The left plots shows that finetuning allows to get rid of the potential hill, leaving only an inflection point suitable for inflation. The right plot shows the non-finetuned generic situation: the potential has two separate critical points and an inflection point in between, thus creating a potential barrier. To move down the throat (towards smaller $\phi$) the inflaton has to cross the barrier and run uphill over a certain interval.}
\end{figure}

Two relevant embeddings that give sizable contributions in the theoretically controllable small $\phi$ regime are the Ouyang \cite{Ouyang:2003df} and the simplest Kuperstein embedding \cite{Kuperstein:2004hy}. They induce terms proportional to $\phi$ and $\phi^{3/2}$ in the potential. Most other embeddings, on the contrary, give rise to contributions proportional to $\phi^p$, where $p>2$, which renders them subleading in the small $\phi$ region. They can thus not help to flatten the inflaton potential. For the Ouyang embedding the corrections to the potential, induced by the threshold corrections to the non-perturbative superpotential, vanish after angular moduli stabilization \cite{Burgess:2006cb}. For the Kuperstein embedding, on the other hand, they remain non-trivial, as we will show. The resulting inflaton potential, $V(\phi)$, in this latter case is portrayed in fig.~\ref{twins}. In general, it possesses a maximum and a minimum plus an inflection point in between, as shown in the right figure. With suitable fine-tuning, displayed in the left figure, it can be arranged that the maximum and minimum coincide with the inflection point, the potential hill at small $\phi$ disappears, and the potential becomes flat enough for inflation.

We then study the cosmological evolution implied by the potential $V(\phi)$. It exhibits a slow-roll inflation phase (in particularly $\eta\simeq0$) only in a small region around the inflection point, where $\eta =0$, and at which $\eta$ changes sign (see figure \ref{tre}). Here the potential switches from concave to convex. The reason that only a small portion of the potential can be flattened is that various terms in $V(\phi)$ have different $\phi$ dependence (see eq.~(\ref{effective})). In a fine tuned case (see the left part of figure \ref{twins}), the inflection point can be made flat and a prolonged stage of slow-roll inflation is induced.

If one wants to end inflation with D3 - anti D3 annihilation, the D3-brane has to go all the way down to the tip of the throat towards $\phi \rightarrow 0$, therefore generically running uphill for a certain interval (only in the fine-tuned case where maximum and minimum coincide with the inflection point, does the potential hill disappear and turn into a flat region). We investigate if overshooting the potential hill is possible or whether the inflaton gets stuck in the minimum due to Hubble friction. We find that overshooting is possible if the minimum exhibits a fairly small positive cosmological constant.

The structure of the paper is as follows: in section \ref{sec:sup} we describe the structure of the effective superpotential and its relation to the D7-brane embedding. Section \ref{sec:eta prob} reviews the D-brane inflation $\eta$-problem, which forms a main issue in this paper, explains the type IIB setup and provides the effective potential for the moduli in the warped conifold background. In section \ref{mini1} we perform a minimization of the potential in the K\"ahler modulus and angular directions for a general D7-brane embedding. In section 5 we apply these general results to the Ouyang embedding and compare it with the results for the Kuperstein embedding. Section 6 investigates inflation in the Kuperstein embedding case and analyzes the cosmological evolution around the minimum of the potential followed by an analysis of the uphill evolution. In section \ref{sec:anti D3} we discuss various forces acting on the D3- and anti D3-brane and comment on their relative importance. We conclude in section 8. A couple of appendices provide further technical details. \ref{app:conifold} collects some details about the warped conifold background. \ref{a:p} lists and discusses the related parameters. \ref{scr dep} analyzes the dependence of the stabilized volume modulus $\sigma_c$ on the uplifting potential and the inflaton. \ref{r3/2} shows that the coefficient of the $\phi^{3/2}$ term in the inflaton potential, for the Kuperstein embedding, is non-positive. This feature determines the general structure of the potential which is derived in \ref{max min}.

\textbf{Note added:} Almost simultaneously with the submission of this paper, two other papers appeared \cite{Baumann:2007np}, \cite{Baumann:2007ah} which address the same issue and come to similar conclusions. The second version of our paper corrected a mistake in the angular minimization of our earlier version. The corrected calculation leads to the conclusion that fine-tuning to generate inflation is possible for the Kuperstein but not the Ouyang embedding in agreement with \cite{Burgess:2006cb,Baumann:2007np,Baumann:2007ah}. We thank James Cline for helpful correspondence.


\section{Superpotential}\label{sec:sup}

The Gukov-Vafa-Witten flux superpotential (GVW) $W_0$ \cite{Gukov:1999ya,Gukov:1999gr} can fix the dilaton and complex structure moduli in type IIB flux compactifications. The K\"ahler moduli, on the other hand, are stabilized by a non-perturbative superpotential $W_{np}$ \cite{Curio:2001qi,Kachru:2003aw}. The latter breaks the no-scale structure because of its explicit K\"ahler moduli dependence.

$W_{np}$ can be generated either by Euclidean brane instantons, D3-instantons in our case, or gaugino condensation on a stack of D7-branes. In type IIB compactifications with a single K\"ahler modulus either of the two effects is sufficient (in general, both effects arise together and lead to economical ways of stabilizing further K\"ahler moduli, as shown recently for heterotic M-theory compactifications \cite{Curio:2006dc}). In both cases, the branes wrap a divisor $\Sigma$ of the Calabi-Yau manifold. The embedding is then specified by a section of a divisor bundle, $f=[\Sigma]$. The D3-brane backreacts on the metric background and hence alters the instanton (gaugino condensation) action. This effect sources an additional interaction for the D3-brane position moduli and can be described by an $f(w)$ dependence of $W_{np}$, where $w$ indicates the position of the D3-brane in the Calabi-Yau manifold (for the  conifold $w$ will be a set of three out of four projective coordinates, see \ref{app:conifold}). In \cite{Baumann:2006th} (see also \cite{Ganor:1998ai}, \cite{Berg:2004ek}, \cite{Giddings:2005ff}, \cite{Koerber:2007xk}) it was obtained that
\be \label{sup}
W_{np} = A(w)e^{-a\rho}
\equiv A_0 f(w)^{1/n} e^{-a\rho} \; ,
\ee
where $A_0$ depends on the already stabilized complex structure moduli. In the sequel, following KKLT \cite{Kachru:2003aw}, we assume that the complex structure moduli have been stabilized by fluxes at a scale hierarchically higher than the scale of inflation (although this might not be the generic case, we are focussing our investigation on a corner of the landscape where this assumption holds). As in this paper we are interested in the dynamics of inflation, we will in the rest of the paper treat $A_0$ as a constant.
In addition, to simplify the analysis, we assume a compactification with a single K\"ahler modulus, which we denote $\rho=\sigma+i b$. Furthermore, $a=2\pi/n$, with $n$ being the number of D7-branes in the stack producing gaugino condensation ($n=1$ for the Euclidean D3-brane case).

By an adequate shift of the axion $b=\mathrm{Im}\rho$, $A_0$ can be taken to be real. The total superpotential therefore is
\be\label{sup tot}
W=W_0+A(w)e^{-a\rho}\,.
\ee
A class of supersymmetric embeddings has been found in \cite{Arean:2004mm}. It is given by
\be \label{susyembed}
f(w) \equiv 1 - \frac{\prod_{i=1}^4 w_i^{p_i}}{\mu^P} = 0\,,
\ee
where $p_i \in \mathbb{Z}$, $P \equiv \sum_{i=1}^4 p_i$, and $\mu
\in \mathbb{C}$ are (constant) parameters defining the embedding of the
D7-branes. The simplest choice of parameters $p_i=\delta_{1,i}$ reproduces the Ouyang embedding \cite{Ouyang:2003df}. The $p_i$ have to be integers by holomorphicity\footnote{In the Ouyang case, the integer $p_1=P$ can be interpreted as the number of times a D7-brane is wrapped around the 4-cycle.}.

Another very simple embedding is the Kuperstein embedding \cite{Kuperstein:2004hy}
\be
f(z)\equiv1-\frac{z_1}{\mu}\,,
\ee
which is expressed in terms of alternative coordinates on the conifold (see appendix \ref{app:conifold}). The $\{z_i\}$ are linear combinations of the $\{w_i\}$, so again only integer powers of $\{w_i\}$ (equivalently $\{z_i\}$) are allowed by holomorphicity. This will play a crucial role in the following.


\section{Warped D-Brane Inflation}
\label{sec:eta prob}

In this section, we review the $\eta$-problem arising for D3-brane inflation in a warped throat driven by brane-anti brane attraction, first pointed out in \cite{Kachru:2003sx}. Then we add threshold corrections to the analysis and obtain the F-term potential that we will study in the next section.

\subsection[The $\eta$-Problem from Volume Stabilization]{The $\eta$-Problem from Volume Stabilization}

It was pointed out in \cite{Kachru:2003sx}, that the strongest force felt by the D3-brane comes from the mixing of open string moduli with the overall volume, once the latter is stabilized \`a la KKLT. To see how this comes about, let us briefly review the KKLT setup \cite{Kachru:2003aw}.  Upon reducing the 10-dimensional type IIB superstring theory over the warped metric background
\be\label{metric}
ds_{10}^2 = \hh^{-1/2} ds_4^2 + \hh^{1/2} ds_6^2 \; ,
\ee
where $h$ is the warp factor in the presence of imaginary self-dual fluxes and orientifold planes, we obtain a 4-dimensional, $N=1$ supergravity theory. First let us assume that no D3-branes are present, we will add it in a second step. Then the prefactor $A$ in eq.~(\ref{sup tot}) is a constant because, as we mentioned, we assume that the complex structure moduli and the dilaton have already been stabilized. The resulting F-term potential for the K\"ahler moduli plus the anti D3-brane uplifting term are \cite{Kachru:2003aw}
\be\label{pott}
V_{\rm dS}&=&V_{\rm AdS}+V_{\rm up}\\
&=&\frac{a A_0 e^{-a \sigma } }{2\sigma^2}\left( \frac{1}{3}\sigma a A_0 e^{-a\sigma}+ A_0 e^{-a \sigma}+W_0 \right)+\frac{D}{(2\scr)^2}\nonumber.
\ee
The values of the GVW superpotential $W_0$ and $D$ (proportional to $\hh^{-1}_0$, the warp factor at the tip of the throat) depend on the fluxes which stabilize the complex structure moduli and the dilaton. Given the large freedom in the choice of fluxes, we will treat these quantities effectively as tunable constants. It is useful to re-express $\{W_0,D\}$ in terms of two other quantities $\{\ssi_0,\beta\}$ as
\be
W_0&=&-A_0e^{-a\ssi_0}\left(1+\frac23 \ssi_0 a\right)\,,
\label{param2} \\
D&=&\frac23\, \beta \, \ssi_0\, a^2|A_0|^2e^{-2a\ssi_0}\,.
\label{paramb}
\ee
The parameters $\{\ssi_0,\beta\}$ have the following meaning: $\ssi_0$ is the KKLT minimum, i.e.~the value of $\ssi$ in the AdS minimum obtained for $D=0$. Adding the uplifting ($D\neq0$), the minimum of eq.~(\ref{pott}) is shifted to $\ssi=\scr$ which is very close to $\ssi_0$ (see appendix \ref{scr dep}); in fact $\scr-\ssi_0\equiv\Delta \ll\ssi_0$. Hence, $\ssi_0$ is an estimate of the position of the actual minimum and if it is chosen to be large enough we can neglect $\al$ corrections. As regards $\beta$, it parameterizes the uplifting in such a way that a Minkowski vacuum corresponds to $\beta\simeq1+2\Delta/\ssi_0$, i.e.~a value slightly larger than one while for $\beta\gtrsim 1+2\Delta /\ssi_0$ we have a dS vacuum (see \ref{scr dep}).
At the minimum $\sigma=\scr$, the potential in eq.~(\ref{pott}) takes the value
\be\label{no D3}
V_{\rm dS}|_{\scr}= -\frac{a^2 |A_0|^2 e^{-2a \scr } }{6\scr}+\frac{D}{(2\scr)^2}\,,
\ee
where we have neglected terms suppressed by $\Delta/\sigma_0$.

Now we want to add to the picture a D3-brane located at a point $w$ in the Calabi-Yau manifold. This induces several modifications to the potential in eq.~(\ref{pott}). The volume of 4-cycles is shifted by a $w$-dependent quantity, i.e.~it acquires a dependence on the position of the D3-brane. In \cite{DeWolfe:2002nn} it was proposed that, in the simple case of a single K\"ahler modulus (determining the overall volume), the K\"ahler potential is
\be \label{kah}
K = -2\mathrm{log}(\V)
= -3\mathrm{log}[\rho+\overline{\rho}-\gamma k(w,\overline w) ] \equiv -3{\rm log}R \,,
\ee
where $\gamma\equiv\kk^2 \scr T_{D3}/3$ is a constant, $T_{D3}$ being the D3-brane tension, and $k(w,\overline w)$ is the K\"ahler potential of the Calabi-Yau manifold evaluated at the position of the D3-brane. The metric of a compact Calabi-Yau threefold is not known, but as long as we are interested in the dynamics inside a warped throat, we can approximate it by the conifold metric. We have in mind a warped deformed conifold (eventually cut and glued to a compact Calabi-Yau manifold as in GKP \cite{Giddings:2001yu}), but we will always consider regions far away from the tip where the metric is well approximated by the singular conifold\footnote{Very close to the tip, where the deformation cannot be neglected, one has $k(w,\overline w)=r^3+\mathrm{const.}$ \cite{Candelas:1989js,DeWolfe:2007hd}. As noticed in \cite{Pajer:2008uy}, this implies that the effect of moduli stabilization very close to the tip of a warped deformed conifold is to generate a term proportional to $\phi^3$ instead of $\phi^2$ as is the case for the singular conifold. This $\phi^3$ term could in principle be canceled by the threshold corrections to $W_{np}$ we are considering here. The cancellation would be, however, only valid very close to the tip, i.e.~for a short range of values of the inflaton field.}. This allows us to use the singular conifold K\"ahler potential $k(w,\overline w)=r^2$, where $r$ is the radial direction of the conifold (see \ref{app:conifold}).

Now we use the K\"ahler potential in eq.~(\ref{kah}) with $k(w,\overline w)=r^2$ to calculate the potential in eq.~(\ref{pott}) which becomes therefore $r$ dependent. As $r$ is our inflaton candidate, it is convenient to express the result in terms of the canonically normalized inflaton field $\phi=r\,\sqrt{T_3}$
\be\label{Vds}
V_{\rm dS}&=&\frac{M^4_{Pl}}{(\phi^2-6M^2_{Pl})^2}\left( \frac{9D}{\scr^2}-\frac{6|A_0|^2a^2e^{-2a\scr} } {\scr} \right) \nonumber \\
&\equiv&3H^2 \,\frac{36M^6_{Pl}} {(\phi^2-6M^2_{Pl} )^2}\,,
\ee
where we have introduced the Hubble parameter $H$ for $\phi=0$ (and neglected $\dot{\phi}^2$). The fields $r$ and $\phi$ have dimension of a length and a mass, respectively. On the other hand, $\sigma$ has been normalized to be dimensionless. The slow-roll parameter $\eta$ is then
\be
\et &=& M^2_{Pl}\frac{V_{\rm dS}''}{V_{\rm dS}} = M^2_{Pl} \left[3H^2\,\frac{144(6M^2_{Pl}+5\phi^2)}{(\phi^2-6M^2_{Pl})^4}\right]
\left[3H^2\,\frac{36}{(\phi^2-6M^2_{Pl})^2}\right]^{-1} \nonumber \\
&=&\frac{4(6M^2_{Pl}+5\phi^2) } {( \phi^2-6M^2_{Pl})^2 }M^2_{Pl} \label{etakklt}\,.
\ee
From its definition, $\phi$ is positive and smaller than $\sqrt{6}M_{Pl}$. At this value in fact, the volume $\V$ in eq.~(\ref{kah}) becomes zero (assuming that the K\"ahler modulus $\rho$ has reached its minimum $\rho+\overline\rho=2\scr$) and the shifted K\"ahler potential becomes singular. Therefore $\et$ is always bigger than $2/3$ (the conformal value attained for $\phi=0$ \cite{Kachru:2003sx}) and slow-roll inflation never takes place without threshold corrections.


\subsection{F-term Potential for the Conifold}

In this subsection we repeat the calculation of the last subsection but now taking into account the threshold corrections to the non-perturbative superpotential discussed in chapter \ref{sec:sup}, i.e.~we allow for a generic $A(w)$. We start with the 4-dimensional, $N=1$ supergravity scalar potential
\be
V_F = e^K \left( K^{\overline b a}D_a W\overline{D_b W}
-3|W|^2 \right) \; .
\ee
The indices $a,b$ run over the complex fields $\rho$ and $w=w_i$ with $i$ running over three of the four homogeneous coordinates $w_A$ introduced in \ref{app:conifold}.



For the K\"ahler potential of eq.~(\ref{kah}) and a generic superpotential $W$ the resulting F-term potential takes the form (cf.~\cite{Burgess:2006cb})
\be \label{kklt}
V_F&=&V_{\mathrm{KKLT}}+\Delta V\\
   V_{\mathrm{KKLT}} &=& \frac{\kappa_4^2}{ 3 R^2}  \left[  (\rho+\overline\rho)|W_\rho|^2 -3(\overline{W} W_\rho + {\rm c.c.} )\right]\\
\Delta V &=&\frac{\kappa_4^2}{ 3 R^2}\left[\frac32
    \left(\overline W_{\overline\rho} \sum_i w_i W_i + {\rm c.c.}\right) \label{deltaV}
   +\frac{1}{\gamma} k^{\overline \jmath i} \overline W_{\overline \jmath} W_i \right] \; ,
\ee
where $W_\rho \equiv \partial_\rho W$, $W_i \equiv \partial_i W$. Note that all terms of type $K^{\overline \jmath i} \overline W_{\overline \jmath}W K_{i}$ cancel out precisely. Thus, $V_F$ would vanish if the superpotential were independent of $\rho$ and $w_i$ because of the no-scale structure. But it is not and it is not. Indeed, using the superpotential in eq.~(\ref{sup tot}), with a generic $A(w)$, one finds
\be \label{expl}
    V_{\mathrm{KKLT}}&=& \frac{\kappa_4^2}{ 3 R^2}\left[ \left[(\rho+\overline\rho)a^2+6a\right]
    |A|^2e^{-a(\rho+\overline\rho)} + 3a(\overline W_0 Ae^{-a\rho} +\mathrm{c.c.})\right]\\
    \Delta V&=&  \frac{\kappa_4^2}{ 3 R^2} \left[   -\frac32a
        \Big(\overline A \sum_i w_i A_i + {\rm c.c.}\Big)
        + \frac{1}{\gamma}  k^{\overline \jmath i}\overline A_{\overline \jmath} A_i \right] e^{-a(\rho+\overline\rho)} \; ,
\ee
where $A_i \equiv \partial_i A$. The separation into two terms is due to the fact that $\Delta V$ is non-vanishing only when $A$ is a non-trivial function of the $w_i$. In the last subsection, where we assumed that $A$ is a constant, $\Delta V$ was absent. Note also that $V_{\mathrm{KKLT}}$ is not the same as $V_{\rm AdS}$ in eq.~(\ref{pott}) of the last subsection. It differs in two ways: first, in $V_{\mathrm{KKLT}}$ there is also a dependence on the angular moduli through the non-constant $A(w_i)$; second, due to the backreaction of the mobile D3-brane the volume modulus has become $R=2\ssi-\gamma r^2$ rather than simply $2\ssi$ and has acquired a dependence on the D3-brane radial position.

Since we want to find out whether warped D3-brane inflation is possible in this setting, we need to be in a dS space. This can be achieved by adding an uplifting term
\be
V_{\rm up} = \frac{D}{R^2}
\ee
to $V_F$. The uplifting breaks supersymmetry and lifts the vacuum to a dS one. For concreteness we will think of this term as coming from the warped anti D3-brane tension \cite{Kachru:2003aw}. This is not essential for our purposes and other upliftings, such as D-term or F-term upliftings can be used as well. Two comments are in order. First, the $R^2$ dependence is appropriate for the warped throat under consideration whereas an ordinary compact six-manifold would generate an $R^3$ dependence instead, as discussed in \cite{Kachru:2003sx}. Second, due to the backreaction of the mobile D3-brane there is a dependence on its position $r$ in the denominator, which uses the corrected volume modulus $R$ rather than $\sigma$.

\section{Critical Points of the Potential} \label{mini1}

Our eventual goal is to identify the inflaton with the mobile D3-brane position modulus $r$ and to study whether its potential
\be
V = V_{\rm KKLT}  + V_{\rm up}  + \Delta V\; ,
\ee
can lead to viable inflation. To this end we have to ensure that there is no steep runaway in some other direction in moduli space. Therefore, next we analyze the stabilization of all moduli besides $r$, which comprises the volume modulus $\sigma$, its axionic partner $b$ and the angular moduli $\theta_1,\theta_2,\phi_1,\phi_2,\psi$. As we want to restrict ourselves to the case of single field inflation we have to require that the D3-brane motion does not modify considerably the stabilization of the other fields. A convenient regime to consider is
\be \label{f=1}
|f(r)|^{1/n} - 1 \ll 1\,,
\ee
\begin{figure}
\parbox{0.5\textwidth}{\includegraphics[height=0.5\textwidth]{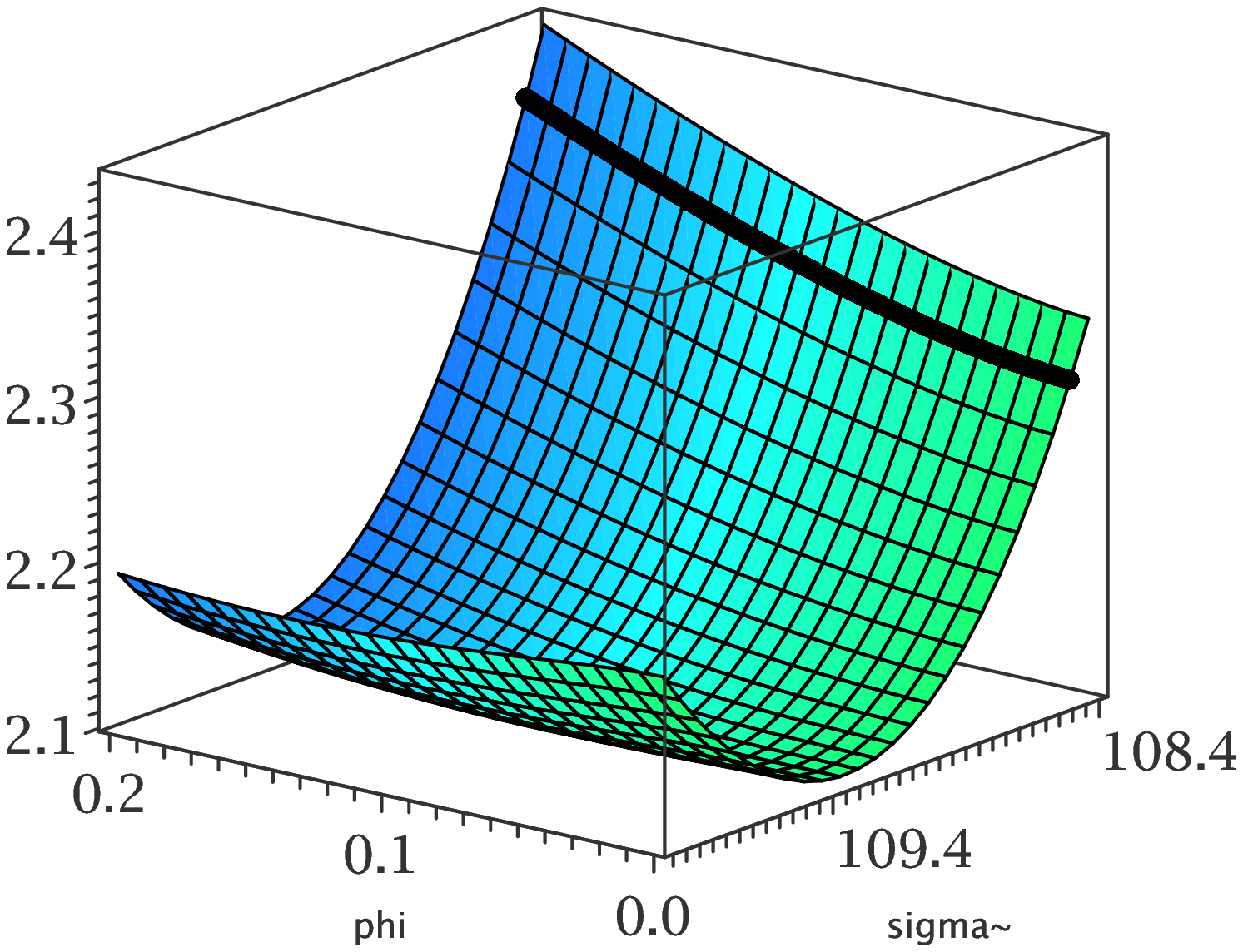}}
\parbox{0.5\textwidth}{\includegraphics[width=0.5\textwidth]{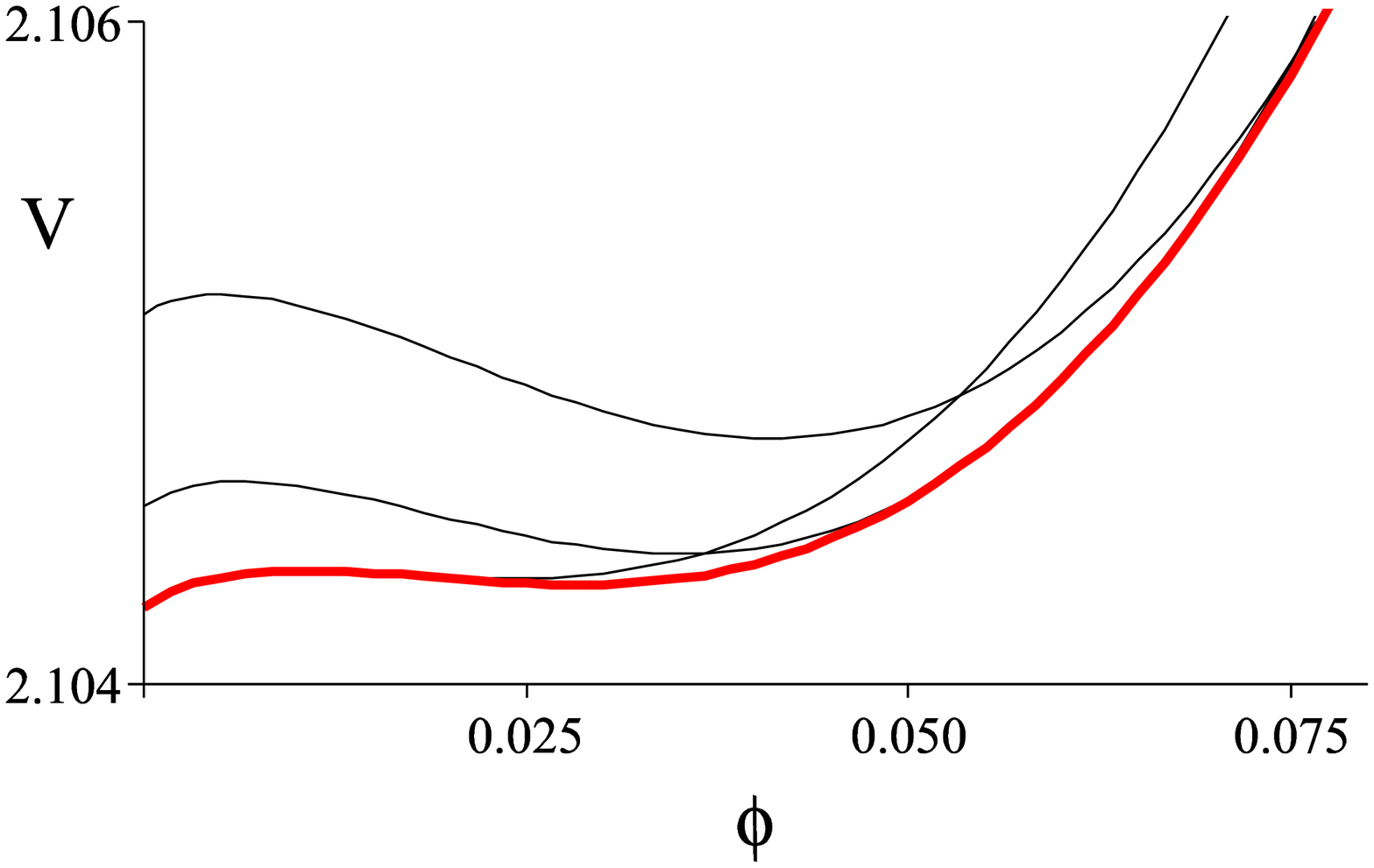}}
\caption{\label{two}On the left: the dependence of the potential on $\phi$ and $\sigma$ near the minimum. The black thick line is the value of $\scr$ one would get neglecting the uplifting term (using just eq.~(\ref{sigma0})). Clearly if one is interested in inflation dynamics, neglecting $V_{\rm up}$ is inconsistent. On the right: the black thin lines are the potential (times $10^{16}$) evaluated for different but $\phi$ independent $\scr$. The red thick line is obtained plotting $V(\phi,\scr(\phi))$ (times $10^{16}$). Again one clearly sees that it is inconsistent to study inflation just in the $\phi$ direction for fixed $\scr$}
\end{figure}
so that the critical value of the volume modulus $\scr$ will change only slightly during the inflationary dynamics (see eq.~(\ref{Del1}) and the related discussion). Although the dependence of $\scr$ on $\phi$ is mild (so that during the inflaton motion the minimization of $\sigma$ is only slightly corrected), it is crucial to determine the correct shape of the effective potential for the inflaton $V(\phi)$ (see also \cite{Baumann:2007ah,Baumann:2007np}). In figure \ref{two} (see also \ref{scr dep}), we compare the effective potential $V(\sigma,\phi)$ for some \textit{fixed} values of $\sigma$, with the correct effective potential $V(\scr(\phi),\phi)$. The sections at constant $\sigma$ of the potential differ even qualitatively from the correct effective potential $V(\scr(\phi),\phi)$. In the following we will work in the regime specified by eq.~(\ref{f=1}).


\subsection{Axion Stabilization}

It is easiest to start the moduli stabilization analysis with the axion field $b$. One observes that it makes its appearance only
in the second term of $V_{\mathrm{KKLT}}$
\be
3a(\overline{W}_0\,A\,e^{-a(\sigma+i b)}+{\rm c.c.})&=&3a|W_0\,A|e^{-a\sigma}(e^{-i(ab-\alpha) } +e^{i(ab-\alpha)}) \nonumber \\
&=&6a|W_0\,A|e^{-a\sigma}\cos(ab-\alpha) \,,
\ee
with $\alpha$ denoting the phase of $\overline{W}_0\,A$. This term acquires its minimum when
\be
b_c = \frac{1}{a}\left[\alpha+(2p-1)\pi \right] \; , \quad
p\in\mathbf{Z} \; ,
\label{AxionFix}
\ee
and turns into minus its absolute value. This fixes the axion and implies for the KKLT part of the potential
\be
V_{\mathrm{KKLT}} = \frac{\kappa_4^2}{3 R^2} \left[ 2a\left(a\sigma+3\right) |A|^2e^{-2a\sigma} - 6a |W_0 A| e^{-a\sigma} \right] \; .
\ee


\subsection{Volume Modulus Stabilization}

This section is devoted to the minimization of the volume $\ssi$, which is more involved than the axion minimization. The reason is, that, with $r$ being our inflaton candidate, it is particularly important to determine the $r$ dependence of the critical value $\ssi_c(r)$ of the modulus $\ssi$.

The criticality condition, $\partial_\sigma V = 0$, which determines $\sigma_c$, reads
\be\label{crit}
(aR_c+2)(V_{\rm KKLT} + \Delta V) + 2V_{\rm up}
= \frac{\kappa_4^2 a^2}{3R_c}|A| e^{-a\sigma_c} \Big( |A| e^{-a\sigma_c} - 3|W_0| \Big)
\, ,
\ee
where $R_c \equiv 2\sigma_c-\gamma r^2$. If $\Delta V$, $V_{\rm up}$ and the mobile D3-brane were absent, such that $R\rightarrow 2\sigma$, the criticality condition would lead to the original KKLT result \cite{Kachru:2003aw}
\be
V_{\rm KKLT,0} = -\frac{\kappa_4^2 a^2 A_0^2 e^{-2a\sigma_0}}{6\sigma_0}
\, ,
\ee
with the KKLT critical volume, $\sigma_c\rightarrow\sigma_0$, defined implicitly by
\be \label{sigma0}
W_0 = - A_0 e^{-a\sigma_0} \left(\frac{2}{3}a\sigma_0  + 1 \right)
\, ,
\label{VolFix}
\ee
where the fixed axion value has been used. Once $V_{\rm up}$ and the mobile D3-brane are added, the critical volume, $\sigma_c$, is shifted away from the constant $\sigma_0$
\be
\sigma_0 \; \stackrel{V_{\rm up}, \, D3}{\longrightarrow} \; \sigma_{c} \,.
\ee
Note that $\sigma_{c}$ depends on $D$ and $r$ while $\sigma_0$ does not. We define
\be
\Delta(D,r) = \sigma_c(W_0,D,r)-\sigma_0 \; .
\ee
In what follows, we will use the parameters $\{\beta,\ssi_0\}$ instead of $\{D,W_0\}$ whose definition has been given in eqs.~(\ref{param2}) and (\ref{paramb}). As we pointed out above, the condition that $V_{\rm up}$ uplifts the AdS minimum to dS is now easily expressed through the requirement that $\beta\gtrsim 1+2\Delta/\ssi_0$ (which is very close to, but not exactly one). In the rest of the paper we assume that this condition is fulfilled and therefore the minimum is dS.


Note that the full $r$ (and $\beta$) dependence of $\sigma_c$ is contained in $\Delta$. To calculate $\Delta$ we expand the criticality condition eq.~(\ref{crit}) in $\Delta/\sigma_0$ and use $a\sigma_0\gg 1$ to simplify the result. We obtain
\be\label{sigma min}
\partial_\sigma V = 0: \quad
\begin{array}{|c|}
 \hline \\[-3mm]
 \;\;
 a\Delta (2|f|^{1/n}-1)
 =   \frac{\beta}{a \sigma_0 }|f|^{-1/n} -(1-|f|^{1/n})
 \;\; \\[2mm]
 \hline
\end{array}
\label{Del1}
\ee
where we keep the leading term and first subleading corrections in $1/\sigma_0$ and $\Delta/\sigma_0$ of eq.~(\ref{crit}). This equation determines explicitly the $r$ dependence of $\Delta$ which arises due to the $r$ dependence of $f$.

Without the D3-brane one would have $f=1$ and thus $\Delta=\beta/a^2\sigma_0$ which in turn reduces to zero in the absence of the uplifting ($\beta=0$) in agreement with the expectations. Importantly, the consistency of our expansion can be verified from eq.~(\ref{sigma min}), taking into account eq.~(\ref{f=1}) and that $a\ssi_0\gg1$ which leads to
\be
\frac{\Delta}{\sigma_0}=\co \left(\frac1{\sigma_0^2},\frac{|f|^{1/n}-1}{\sigma_0} \right)\ll 1\,,
\ee
and therefore $\Delta\ll\ssi_0\sim\scr$. Note that in general $\Delta$ depends, via the embedding $f$, also on the angular variables $\theta_1, \theta_2, \phi_1, \phi_2, \psi$ whose stabilization we are analyzing next.


\subsection{Angular Moduli Stabilization}

For sake of brevity, let us denote the angular moduli
\be
\theta_1, \theta_2, \phi_1, \phi_2, \psi
\ee
as $\vartheta_\alpha$, $\alpha=1,\dots,5$ and abbreviate $\partial_\alpha \equiv \partial_{\vartheta_\alpha}$. The criticality condition for the angular moduli, $\partial_\alpha V = \partial_\alpha V_F = 0$, does not involve $V_{\rm up}$ which is independent of $\vartheta_\alpha$. The full angular criticality condition thus reads
\be\label{ang mass}
&\phantom{=}& 2\left( (2a^2\sigma_c+6a)|A| - 6a|W_0|e^{a\sigma_c} \right) \partial_\alpha |A| \nonumber \\
&=& \frac{3}{2}a\partial_\alpha \Big(\overline A \sum_i w_i A_i+ {\rm c.c.}\Big)
- \frac{1}{\gamma} \partial_\alpha \big( k^{\overline\jmath i} {\overline A}_{\overline\jmath} A_i \big) \; ,
\ee
where the left-hand side of the equality stems from $V_{\rm KKLT}$ while the right-hand side originates from $\Delta V$.

As we did in the previous section, we replace $\sigma_c$ by $\sigma_0+\Delta$ and expand in $\Delta/\sigma_0\ll1$. Using eq.~(\ref{sigma0}) to evaluate the left-hand side of eq.~(\ref{ang mass}), one can see that the right-hand side of the criticality condition is suppressed by a factor $1/\sigma_0$ and thus does not contribute at leading order. One finds
\be
\sigma_0(2-|f|^{1/n})\partial_\alpha |A| = 0 \; .
\ee
at leading order in $1/\sigma_0$ and $\Delta/\sigma_0$. In view of eq.~(\ref{f=1}), the values of the angular open string moduli that extremize the scalar potential are solutions of
\be
\partial_\alpha V = 0: \quad
\begin{array}{|c|}
  \hline \\[-3mm]
  \;\; \partial_\alpha |f| = 0 \;\; \\[1mm]
  \hline
\end{array}
\label{AngFix}
\ee
These five equations will fix generically all five angular moduli unless the embedding allows for isometries. However, isometries are incompatible with the bulk Calabi-Yau compactification and hence should be broken. For a detailed discussion of this issue see \cite{DeWolfe:2007hd}.


The fixing of the angular moduli leads to
\be
\Delta V
= \frac{\kappa_4^2 |A_0|^2}{12n^2\gamma}
|f|^{-2+2/n} \partial_r |f|
\left( -8\pi \gamma r |f| + \partial_r |f|
\right) \frac{e^{-2a\sigma_c}}{R_c^2} \,,
\ee
while the other two contributions to the potential become
\be
V_{\rm KKLT} = \frac{2\kappa_4^2 a |A_0|^2}{3}
|f|^{1/n} \Big( |f|^{1/n}(a\sigma_c+3) - (2a\sigma_0+3)e^{a\Delta} \Big) \frac{e^{-2a\sigma_c}}{R_c^2}\,,\\
V_{\rm up} = \frac{D}{ R_c^2} \; ,
\ee
where we have used eq.~(\ref{VolFix}) to eliminate $|W_0|$. We have thus achieved a stabilization of all moduli, except for $r$, the inflaton candidate. The dependence of the full potential on $r$ arises from the $r$ dependences of $\sigma_c(r)$, $\Delta(r)$ and $f(r)$.


\subsection{Potential with Moduli Fixed}

We will now study the potential in the large $\sigma_0$ regime for a general embedding $f$. For this we expand the potential in $\Delta/\sigma_0$ and $1/\sigma_0$ and obtain at leading order
\be
V_{\rm KKLT} &=& V_{\rm KKLT,0} |f|^{1/n} (2-|f|^{1/n})
\left[ 1+\frac{\gamma r^2}{\sigma_0}\right] \nonumber \\
\Delta V &=& \frac{V_{\rm KKLT,0}}{32\pi^2\gamma\sigma_0}
|f|^{-2+2/n} \partial_r|f|
\Big[ 8\pi \gamma r|f| - \partial_r|f| \Big]
\label{pot tot} \\
V_{\rm up} &=& \frac{ D}{4 \sigma_0^2}
\left[ 1+\frac{\gamma r^2}{\sigma_0}-\frac{2\Delta}{\sigma_0} \right] \; \nonumber.
\ee
In the expression for $V_{\rm KKLT}$ there are also two terms proportional to $1-|f|^{1/n}$ at order $\co(1/\sigma_0)$. Using eq.~(\ref{f=1}), a posteriori justified in eq.~(\ref{Del1}), we have omitted these terms. We see, using eq.~(\ref{param2}) and eq.~(\ref{paramb}), that $V_{\rm up}$ appears volume suppressed compared to $V_{\rm KKLT}$ and  $\Delta V$ by an additional factor $1/\sigma_0$.

Notice that in eq.~(\ref{pot tot}) we neglect the Coulomb (plus gravitational) attraction between the D3- and the anti D3-brane. The reason is that the Coulomb attraction is very weak due to the warping. This was, in fact, the basis of the KKLMMT proposal \cite{Kachru:2003sx} to achieve slow-roll brane inflation. As we discussed in section \ref{sec:eta prob}, and is seen here explicitly, there are, however, also effects coming from moduli stabilization that render the F-term potential generically steep ($\et\geq2/3$). Our effort will be to make the F-term, eq.~(\ref{pot tot}), flat enough for slow-roll; once this is achieved, we can add the Coulomb potential as well, and study the resulting slow-roll inflation\footnote{Actually, motivated by this hierarchy of importance, one can even go further and omit any anti D3-brane to begin with. The inflaton potential comes then just from the F-term, which is always present. This idea is realized in the model of inflation at the tip constructed in \cite{Pajer:2008uy}.}.

Our analysis of the inflationary dynamics will be based on two assumptions. The first is that $\ssi$ reaches its $\phi$-dependent minimum (given by eq.~(\ref{sigma min})) instantaneously during the inflaton motion. In other words, the system evolves along the $\scr(\phi)$ trajectory in the $\{\ssi,\phi\}$ plane. For this assumption to be satisfied, the $\ssi$ direction should always be much steeper than the $\phi$ direction which is actually the case here (see also the adiabatic approximation of \cite{Baumann:2007ah}).

The second assumption is more subtle and regards the angular directions. We are assuming that the initial conditions of inflation are such that these directions start at their minima. For the Ouyang and Kuperstein embedding that we consider in the next section, the minimum in the angular directions does not depend on the radial position. Therefore if at the beginning of inflation the system is at an angular minimum, it will stay there forever. This ad hoc assumption about the initial conditions is ubiquitous in the string inflationary literature and is more a technical than a conceptual issue: if the angular directions are steeper than the radial one, then even if they are excited at the beginning, they will relax in a short time; if they are flatter or comparably steep, then they should be included in the inflationary analysis which would become multi-field in nature. In the latter case one is obliged to rely on numerical methods loosing the intuition that the analytical single-field approach usually gives.

%


\section{Explicit Examples: Ouyang vs Kuperstein Embedding}

In this section we study two explicit supersymmetric D7-brane embeddings into the conifold. They have been discovered by Ouyang in \cite{Ouyang:2003df} and by Kuperstein in \cite{Kuperstein:2004hy}. For the Ouyang embedding, we will find that $\Delta V$ vanishes at the minimum of the angular directions, where $\theta_1=\theta_2=0$. This was first noticed in \cite{Burgess:2006cb}. As a result $\widetilde \psi$, defined by
\be\label{psi def}
\ps = \frac{1}{2}(\psi-\phi_1-\phi_2)\,,
\ee
remains unfixed. For the Kuperstein embedding, on the other hand, $\Delta V$ does not vanish at the minimum of the angular directions and can modify $\eta_{KKLT}\simeq2/3$ (see also \cite{Baumann:2007ah,Baumann:2007np}). It is worth noticing that, in the Ouyang case, if the maxima in the angular directions are inserted in $\Delta V$ then the resulting effective potential $V(\phi)$ is exactly the same as in the Kuperstein case. Of course this radial trajectory (that we will analyze in section \ref{infl}) is physically interesting only in the Kuperstein case, where it is stable in the angular directions (an exhaustive analysis of this issue, with a detailed calculation has been given in \cite{Baumann:2007ah}).


\subsection{Ouyang Embedding}

The Ouyang embedding \cite{Ouyang:2003df} is defined by the zeros of
\be\label{e:O}
f(w_i) = 1-\frac{w_1}{\mu}\,.
\ee
Using eq.~(\ref{w}), one derives
\be
|f|^2 = 1 - 2 \frac{r^{3/2}}{|\mu|} \sin\frac{\theta_1}{2} \sin\frac{\theta_2}{2} \cos\ps
+ \frac{r^3}{|\mu|^2} \sin^2\frac{\theta_1}{2} \sin^2\frac{\theta_2}{2} \; ,
\ee
We will take $\mu$ to be real and positive because a possible phase can be absorbed in a shift of $\ps$. The two directions perpendicular to $\ps=\textrm{const.}$ are at this point exactly flat. They will eventually get a mass but their explicit value does not affect the effective potential for the inflaton.

The system of equations fixing the angles, $\partial_\alpha|f|=0$,  turns into
\be
\theta_1 : & -\frac{r^{3/2}}{\mu} \cos\frac{\theta_1}{2} \sin\frac{\theta_2}{2} \cos\ps
+\frac{r^3}{\mu^2} \sin\frac{\theta_1}{2} \cos\frac{\theta_1}{2}
\sin^2\frac{\theta_2}{2} = 0 \\
\theta_2 : & -\frac{r^{3/2}}{\mu} \sin\frac{\theta_1}{2} \cos\frac{\theta_2}{2} \cos\ps
+\frac{r^3}{\mu^2} \sin^2\frac{\theta_1}{2} \sin\frac{\theta_2}{2} \cos\frac{\theta_2}{2} = 0 \\
\phi_1,\phi_2,\psi : & \;\;\;\;\, \frac{r^{3/2}}{\mu} \sin\frac{\theta_1}{2} \sin\frac{\theta_2}{2} \sin\ps = 0 \; .
\ee
This system of equations has two kinds of solutions (angular critical points)
\be
\theta_1 &=& \theta_2 = \pi\,, \qquad \ps = 0, \pi \label{psi fixed} \\
\theta_1 &=& \theta_2 = 0 \; \textrm{ and } \; \ps \; \textrm{ unfixed }
\ee
A detailed study \cite{Burgess:2006cb} (see also \cite{Baumann:2007ah}) of the Hessian matrix shows that the solution corresponding to a minimum is $\theta_1 = \theta_2 = 0$. Here we notice that the other angular direction $\ps$ is not flat when $\theta_i\neq0$; once we evaluate the potential, however, at $\theta_1 = \theta_2 = 0$, no dependence on $\ps$ remains. The actual value of $\ps$ does not affect the following result. In fact, we get
\be
A &=& A_0 f^{1/n} = A_0 \,, \nonumber\\
\Delta V &=& 0\,,
\ee
so that the potential is exactly $V_{KKLT,0}$, leading to $\eta\simeq2/3$. In this case no fine tuning is possible \cite{Burgess:2006cb}.
The other extremum, $\theta_1 = \theta_2 = \pi$, corresponds to a maximum. In this case $\ps$ is fixed (see eq.~(\ref{psi fixed})) but not the two perpendicular directions in $\{\phi_1,\phi_2,\psi \}$ space. If one substitutes these angular values (corresponding to the maximum), one finds
\be
A &=& A_0f^{1/n} = A_0\left( 1+\frac{r^{3/2}}{\mu}\right)^{1/n}\simeq A_0\left(1+\frac{r^{3/2}}{\mu n}\right)\,, \\
\Delta V &=& \frac{\kappa_4^2|A|^2e^{-2a\sigma}}{n^2 R^2}\left[ \frac{2\pi r^{3/2}}{\sqrt{2}\mu+r^{3/2}}+\frac{r}{\gamma (\sqrt{2}\mu+r^{3/2})^2} \right] \; ,
\ee
where in the last step we have used eq.~(\ref{f=1}), that here translates into $r^{3/2}\ll \mu$ and implies that the D3-brane is located further down in the throat than the D7-brane (which extends down to $r_{D7}^{3/2}=\mu$). As we will see in the next section (see also \cite{Baumann:2007ah}), the effective potential $V(\phi)$ that one obtains using this maximum (unstable in the angular directions) is exactly the same as the one for the Kuperstein embedding in eq.~(\ref{effec}), with the angular moduli being at a minimum.


\subsection{Kuperstein Embedding}

The Kuperstein embedding \cite{Kuperstein:2004hy} is defined by the zeros of
\be\label{e:K}
f(z)=1-\frac{z_1}{\mu}\,,
\ee
where now we parameterize the conifold with alternative coordinates $\{z_i\}$ (see \ref{app:conifold}, in particular eq.~(\ref{z})). This embedding has no directions along which $\Delta V=0$ \cite{Baumann:2007ah,Baumann:2007np}. Two trajectories extremize the potential in the angular directions: $z_1=\pm r^{3/2}/\sqrt{2}$, but only the one with the negative sign corresponds to a minimum. The correction to the potential then becomes \cite{Baumann:2007ah,Baumann:2007np}:
\be \label{effec}
\Delta V&=&\frac{\kappa_4^2|A|^2e^{-2a\sigma}}{n^2 R^2}\left[ - 2\pi \mathrm{Re} \frac{z_1} {\mu-z_1}+\frac{r}{\gamma |\mu-z_1|^2}\left(1-\frac{|z_1|^2}{2r^3} \right) \right] \nonumber \\
&=&\frac{\kappa_4^2|A|^2e^{-2a\sigma}}{n^2 R^2}\left[ \frac{2\pi r^{3/2}}{\sqrt{2}\mu+r^{3/2}} + \frac{r}{\gamma (\sqrt{2}\mu+r^{3/2})^2} \right]
\ee
which is exactly the same as in the Ouyang case after choosing the (unstable) trajectory $w_1=-r^{3/2}$. The fact that the minus sign corresponds to the stable trajectory ($z_1=-r^{3/2}/\sqrt{2}$) is crucial for the fine tuning of $\eta$. Indeed it determines that the correction to $\eta_{KKLT}\simeq 2/3$ comes with a minus sign and a cancellation is possible.

The potential we have written still depends on $\sigma$. To obtain the effective potential for the inflaton we need to extremize the potential with respect to $\sigma$, i.e.~use eq.~(\ref{sigma min}). The minimization of the volume can be straightforwardly carried out numerically. An analytical estimate is given by (see \ref{scr dep} for its derivation)
\be\label{traj}
\sigma_c=\sigma_0+\frac{\beta}{a^2\sigma_0}+\frac{r^{3/2}}{an\mu}
+\dots\,,
\ee
where the ellipsis stands for terms suppressed by higher powers in $r^{3/2}/\mu$ or $1/\sigma_0$. We use this expression for the $r$-dependent critical value of $\sigma$ to transform the potential $V(\sigma,r)$ into a potential for a single field $V(r)=V(\sigma_c(r),r)$. This implicitly assumes that the dynamics in the $\sigma$ direction is much faster than in the $r$ direction such that the evolution of the system is well approximated by the trajectory $\sigma_c(r)$ in the $(\sigma,r)$ space. Eventually, the effective potential has to be expressed in terms of the canonically normalized field $\phi$.


\section{Inflation} \label{infl}

In the previous sections we calculated the potential for the radial position $r$ of the D3-brane in the throat, once all other fields have reached their minimum value. In this chapter we investigate if the potential we have obtained can provide phenomenologically viable inflation.

The first step is to rewrite the potential in terms of a canonically normalized field (to which we will refer in the following as the inflaton)
\be
\phi=\sqrt{T_{D3}}   \, r\,.
\ee
We notice that $r$ has the dimension of a length while $\phi$ has that of a mass, as it should be for a canonically normalized scalar in four dimensions.

As we have seen in section \ref{sec:eta prob}, $V_{\rm KKLT,0}$ depends on the inflaton as
\be
V_{\rm KKLT, 0}= 3H^2 \frac{36M_{Pl}^6}{(\phi^2-6M_{Pl}^2)^2}\simeq 3 H^2 M_{Pl}^2+H^2 \phi^2+\dots
\ee
for small $\phi$. This prevents slow roll because
\be
\eta=M^2_{Pl}\frac{V''}{V} \gtrsim \frac23\,.
\ee
If we want to have a flat potential, we thus need another term of the same size but opposite sign that we can fine tune to cancel the $2/3$. The new terms in the potential, eq.~(\ref{pot tot}), coming from the dependence of the non-perturbative superpotential on $\phi$, are proportional to $|f|^{1/n}$ or to $\phi |f|^{1/n}$. The known supersymmetric embeddings all depend on integer powers of $w_i\propto \phi^{3/2}$. This, in particular, implies that there is no term, in the small $\phi$ expansion, that can exactly cancel the $\phi^2$ from $V_{\rm KKLT,0}$. The absence of fractional power embeddings $f\propto w_i^p$ with $p$
non-integer might be traced back to the holomorphicity of $f(w_i)$; it seems therefore hard to circumvent this problem.

Also, all those embeddings for which $f\propto 1+w_i^p$ with $p > 1$ vanish much faster than
$V_{\rm KKLT,0}$ for $\phi \rightarrow 0$ and do not help to flatten the potential. From this observation, it follows that embeddings of the ACR family \cite{Arean:2004mm} with $p>1$ are not helpful to cancel the  $\eta_{KKLMMT}\simeq2/3$, at least for small $r$. Further study is needed to see if there is a region where $r$ is large enough so that the effects of higher ACR embeddings become relevant, and at the same time, that region might still be well described by the conifold geometry (i.e.~before the cut of the conifold and the gluing to the Calabi-Yau manifold become relevant).

Two embeddings that produce corrections to the scalar potential proportional to $\phi$ and $\phi^{3/2}$ (as opposed to $\phi^p$ with $p>2$) are the Ouyang (which is as well in the ACR family with $p=1$) and the Kuperstein embedding. For the former, once the angular minimization is performed, the corrections to the scalar potential vanish \cite{Burgess:2006cb}. For the latter this is not the case and the potential is indeed modified, as shown in eq.~(\ref{effec}) \cite{Baumann:2007np,Baumann:2007ah}.


\subsection{The Effective Inflaton Potential}

\begin{figure}
\centering
\includegraphics[height =0.5 \textwidth, width=0.7 \textwidth]{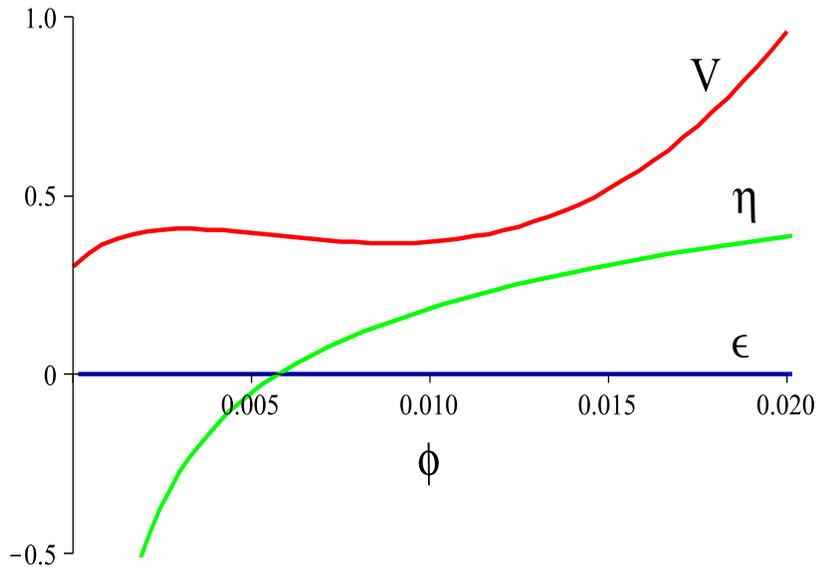}
\caption{\label{tre}The plot shows the potential $V(\phi)$ (red) and the slow-roll parameters $\eta(\phi)$ (blue) and $\epsilon(\phi)$ (black). The latter is so small that it can hardly be distinguished from the $\phi$ axis. Next to the tip of the throat the potential has generically a maximum and a minimum. For $\phi$ large enough the potential grows like $\phi^2$ and $\eta$ is of order one (or bigger). But for $\phi\rightarrow 0$ the curvature of the potential changes at the inflection point and $\eta$ switches sign (and eventually diverges at $\phi=0$).}
\end{figure}
Considering the region deep inside the throat, we expand the potential for small $r$ resp.~$\phi$, keeping terms up to order $r^2$ resp.~$\phi^2$. These are the most interesting terms since higher terms cannot cancel the $\eta_{KKLT}\simeq 2/3$ from $V_{KKLT,0}$. The result is
\be \label{r exp}
\Vds & \equiv& \Vk + V_{\mathrm{up}}  \\
&=& \Vds^{(0)}+\Vds^{(3/2)}\frac{r^{3/2}}{\mu n}+\Vds^{(2)}\frac{\gamma r^2}{\scr}+\dots\,, \nonumber\\
\Delta V&=&\Delta V^{(1)}r + \Delta V^{(3/2)}\frac{r^{3/2}}{\mu n}+\dots\,\nonumber.
\ee
As shown in \ref{r3/2}, $\Vds^{(3/2)}+\Delta V^{(3/2)} < 0$, so that the $r^{3/2}$ term is always negative. In terms of the canonically normalized field $\phi$, we therefore want to study the effective Lagrangian
\be \label{effective}
\mathcal{L}&=&-\frac{1}{2}\partial_{\mu}\phi\partial^{\mu}\phi -V(\phi) \,, \\
V(\phi)&=&\Lambda+ C_1\phi-C_{3/2}\phi^{3/2}+C_2 \phi^2 \nonumber \,,
\ee
where we have approximated the DBI kinetic term by the canonical one. A comment about this approximation is in order. As long as the kinetic energy of the inflaton $\dot\phi^2$ is small compared to the warped D3-brane tension, the higher terms in the expansion of the square root in the DBI action are negligible and the canonical kinetic term gives a good approximation. In the present setup this condition is easily satisfied because inflation takes place in the middle of the throat where the warping is much weaker than at the tip. In models where inflation takes place close to or at the tip, e.g.~\cite{Chen:2004gc} and \cite{Pajer:2008uy}, the effects of the DBI kinetic term become quickly relevant and can give rise to an interesting phenomenology.

The value of the effective cosmological constant term $\Lambda$ depends on several parameters. The stringy parameters are the 3-form fluxes on the conifold, which determine the stabilization of the complex structure and the dilaton. The problem of how the cosmological constant arises from string theory and which is its most probable value, is outside the scope of the present work (see e.g.~the seminal paper \cite{Bousso:2000xa}). In the following, we will simply consider $\Lambda$ as a free parameter. The coefficients $C_1, C_{3/2}$ and $C_{2}$ are such that the potential always has a maximum and a minimum (see \ref{max min}); an extremal case is when these coincide and one gets a flat inflection point. In \ref{max min} we show how varying the uplifting parameter $\beta$ changes the discriminant (see e.g.~figure \ref{infle}). When the discriminant vanishes, maximum and minimum coincide as displayed in the left part of figure \ref{twins}.

In figure \ref{tre}, we plot $\eta$ together with the potential $V(\phi)$. The slow-roll parameter $\eta$ is small only in a narrow interval around the inflection point, $\phi_{\eta=0}$, at which $V''$ (and therefore $\eta$) vanishes. For $\phi>\phi_{\eta=0}$, $\eta$ becomes of order one or bigger and is dominated by $ V_{\rm KKLT}$. For $\phi<\phi_{\eta=0}$, $\eta$ is instead dominated by the correction $\Delta V$, is negative and diverges when $\phi \rightarrow 0$ (but the potential can not be trusted all the way down to $\phi=0$ because of the deformation of the conifold at $r^{3/2}=\epsilon$ where the anti D3-brane sits and Coulomb and tachyon potentials become relevant). We want to stress that to get $\eta=0$ for some value of $\phi$ does not require any fine tuning. In fact $\eta$ is positive for large $\phi$ and negative for small $\phi$, so that by continuity it has to pass through zero.

A generic initial condition would be to start somewhere inside the Calabi-Yau manifold and then ``fall down'' the throat. We therefore start at some $\phi_{i}$ and slide down towards smaller $\phi$. An important quantitative question is if one can get enough e-foldings and an almost scale invariant spectrum. We study the former question in the next section. An important qualitative question is if one can reach the tip or whether the D3-brane gets stuck somewhere before, preventing reheating via brane antibrane annihilation; we address this overshooting problem in sections \ref{damped} and \ref{s:u}.


\subsection{Inflation through an Inflection Point}

As we show in \ref{max min}, the effective potential always exhibits a maximum and a minimum. These coincide for a particular critical value of the uplifting parameter $\beta$, giving rise to a flat inflection point at some $\phi=\overline{\phi}$. The critical $\beta$ can be estimated analytically from the zero of the discriminant given in eq.~(\ref{discri}). In this section we comment on this fine tuned case. A crucial point is that around $\overline{\phi}$ the Coulomb potential (that we have argued could be neglected in the precedent discussion) has to be taken into account.

For example, consider a potential of the type $V'''(\overline\phi)(\phi-\overline\phi)^3+\Lambda$ , where the interesting case for us is when $\Lambda\gg\phi^3$. The first derivative at the inflection point $\phi=\overline\phi$ is strictly vanishing. As a consequence the slow-roll attractor describes an inflaton that slows down exponentially fast and never reaches the inflection point. There are two effects that regularize this divergence: one is induced by corrections to the strict slow-roll approximation, such as an initial non-slow roll $\dot\phi$. This could allow to pass the inflection point in finite time. A second effect (which co-exists with the first) is induced by the subleading terms that we neglected in the potential; they can have a non-vanishing first derivative at the   inflection point. An example is the generally subleading Coulomb
potential, $V_{D3\overline{D3}}$, that becomes important around $\overline\phi$, where the potential is otherwise flat. Linearly approximating $V_{D3\overline{D3}}$, one gets a potential of the type
\be
V(\phi)\sim\Lambda+ V'(\overline \phi)(\phi-\overline\phi)+V'''(\overline\phi)(\phi-\overline\phi)^3\,.
\ee
In this case, the inflection point is always reached and in fact overshot. The number of e-foldings, $N_e$, that results from the inflationary dynamics is therefore controlled by the value of the first derivative of the potential at $\overline\phi$, $V'(\overline\phi)$. Varying $V'(\overline\phi)$ continuously, from positive to negative values, corresponds to a change of $N_e$ from a few to infinity. An analytical estimate (neglecting corrections to slow roll) gives $N_e\propto 1/V'(\overline\phi)$ for positive $V'(\overline\phi)$ (see e.g.~\cite{Allahverdi:2006we}); for negative $V'(\overline\phi)$ a minimum of $V$ is formed and the issue of overshooting and slow-roll corrections becomes important. In any case, it is clear that an arbitrary large amount of inflation can be obtained by the potential we have calculated, provided that one can fine tune the string theory parameters to obtain a small $V'(\overline\phi)$. Taking into account the effects of the DBI action can only increase the number of e-foldings. In the next two sections we will go beyond the slow-roll approximation and address the question, when does the D3-brane reach the tip and when does it get stuck somewhere in the middle?

A final comment is in order. In the present model the shape of the potential is determined by the F-term, while the Coulomb potential gives only relatively small corrections. These corrections are relevant only in the fine tuned case of a flat inflection point. Even in this case, the Coulomb potential dominates only around the inflection point where the force exerted by the F-term is vanishing. We want to contrast this situation with another expectation which is often found in the phenomenological brane inflationary literature (see e.g.~\cite{Bean:2007eh}). One could have wished to find a way to completely get rid of the F-term effects and have an inflationary model based on a Coulomb potential of the type
\be
V\sim V_0\left(1-\frac{1}{\phi^4}\right)\,,
\ee
which has been widely studied (see e.g.~\cite{Bean:2007eh}). The result of the present investigation is that this is even harder to achieve than expected. For the class of embeddings which we studied, not even fine tuning allows us to cancel completely the F-term effects. We consider this as an indication that in a generic model of brane inflation, the Coulomb attraction is superfluous because the inflaton potential is determined by the F-term potential\footnote{In \cite{Pajer:2008uy}, proceeding along the lines of this argument, a model was constructed where the inflaton potential comes exclusively from the F-term and no anti D3-branes were needed.}.


\subsection{Damped oscillatory phase}\label{damped}

In this and the next section we study the problem of overshooting the potential barrier. We make the following simplifying assumptions: we consider a homogeneous and isotropic universe so that the 4-dimensional Einstein equations reduce to the Friedmann equations; furthermore we assume that all fields have been stabilized except for the inflaton, as shown in the previous sections; finally we neglect effects of the DBI action and approximate it by a canonical kinetic term as discussed around eq.~(\ref{effective}).

\begin{figure}
\centering
 \includegraphics[height=0.5\textwidth]{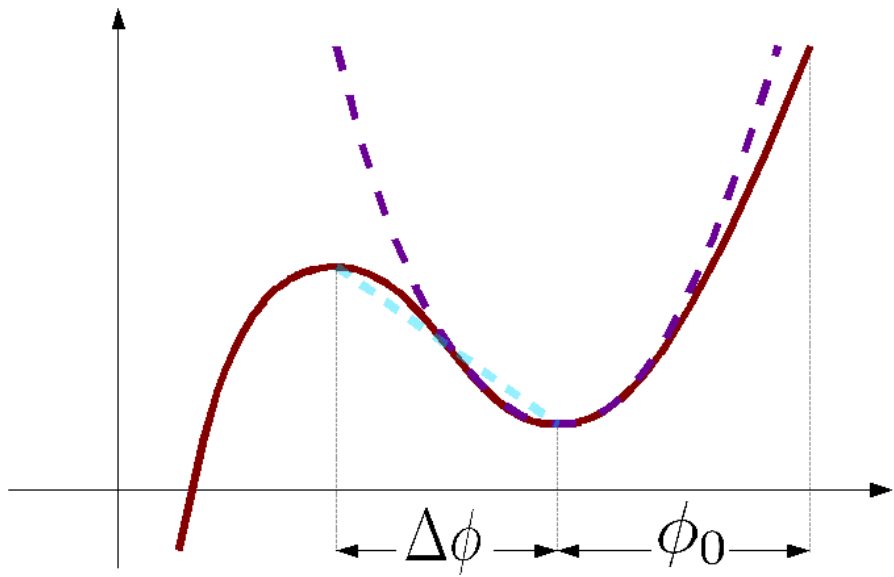}
\caption{The figure summarizes our overshoot analysis. The continuous line is the actual potential, the darker dashed line refers to the discussion of the damped oscillatory phase and the lighter dashed line refers to the uphill phase. \label{fig9}}
\end{figure}

Let us start considering the following potential
\be
V=\Lambda+\frac12 m^2 (\phi-\phi_{\mathrm{min}})^2
\ee
in the vicinity of its minimum, $\phi_{\rm min}$, where $\Lambda \gg m^2 (\phi-\phi_{\mathrm{min}})^2$. The first Friedmann equation gives
\be
H^2=\frac1{3M_{Pl}^2}\left[\Lambda+\frac12 m^2 (\phi-\phi_{\mathrm{min}})^2+\frac12 \dot{\phi}^2\right]\simeq\frac{\Lambda}{3M_{Pl}^2}\,. \label{friedmann}
\ee
Therefore, the equation of motion can be approximated by
\be
\ddot{\phi}+\frac{\sqrt{3\Lambda}}{M_{Pl}}\dot{\phi}
+ m^2(\phi-\phi_{\mathrm{min}})=0\,,
\ee
where the consistency of neglecting $\dot{\phi}^2$ in eq.~(\ref{friedmann}) will be checked at the end. The equation of motion is the same as for a harmonic oscillator with friction. There are consequently three types of solutions:
\begin{itemize}
\item Underdamped: $M_{Pl}^2 m^2>3\Lambda/4$; only in this case the field oscillates around the minimum. The amplitude decreases exponentially on a time scale $\sqrt{3\Lambda}/2M_{Pl}$. If the field starts at $\phi_{i}$ at $t=0$ with $\dot{\phi}_i=0$, we can estimate the speed when it passes the first time through the minimum\footnote{Notice that the formula we give is valid \textit{only} at $t=t_{\mathrm{min}}$ and not for a generic time $t$.} at $t=\tm$ as
\be \label{phidot in}
\dot{\phi}_{\mathrm{min}}=\phi_0 e^{-\sqrt{3\Lambda}\tm/2M_{Pl}}\frac{2m^2M_{Pl}}{\sqrt{3\Lambda}} \cos(\tm\omega)\,,
\ee
where $\omega^2=m^2-3\Lambda/4M_{Pl}^2$, and $\phi_0\equiv\phi_i-\phi_{\mathrm{min}}$ is the distance from the starting point to the minimum (see figure \ref{fig9}).
\item Critically damped and overdamped: $M_{Pl}^2 m^2\leqslant 3\Lambda/4$. There are no oscillations and the field takes an infinite amount of time to reach the minimum (where $\dot{\phi}=0$).
\end{itemize}
The underdamping condition can be rewritten in terms of the slow-roll parameter as $M_{Pl}^2m^2/\Lambda = \eta > 3/4$, where $\eta=\eta(\phi_{\mathrm{min}})$. A rough estimate gives $|\dot{\phi}_{\mathrm{min}}|\sim\eta\phi_0\sqrt{3\Lambda}/M_{Pl}$. Therefore neglecting the kinetic term in the Friedmann equation (\ref{friedmann}) and using $H^2\simeq\Lambda/3M^2_{Pl}$ is legitimate as long as $\eta^2\phi_i^2\ll M^2_{Pl}$.

We want to apply this analysis to our potential eq.~(\ref{effective}) in the case this exhibits a minimum at $\phi=\phi_{\mathrm{min}}$. This could be the case for example if string theory does not allow for an arbitrary fine tuning of the effective parameters $C_1$, $C_{3/2}$ and $C_2$. Around $\phi_{\mathrm{min}}$, the potential is approximated by a harmonic oscillator $V\simeq V''(\phi-\phi_{\mathrm{min}})^2/2$ (see figure \ref{fig9}). We conclude that if $\eta(\phi_{\mathrm{min}})<3/4$ the inflaton reaches the minimum $\phi_{\mathrm{min}}$ only asymptotically in infinite time. There is no graceful exit from inflation as there is no brane annihilation nor (damped) oscillations. The exponential expansion (with cosmological constant $V(\phi_{\mathrm{min}})$) continues forever. On the contrary, if $\eta(\phi_{\mathrm{min}})>3/4$, we find at the minimum that $\dot{\phi}_{\mathrm{min}}\neq0$. This allows for the possibility to climb up to the maximum of the potential, overshoot it and reach the tip of the throat where annihilation with the anti D3-brane will take place.


\subsection{Uphill inflation}\label{s:u}

In this section we study what happens when the inflaton rolls uphill. We will use our results to address the issue of overshooting the potential barrier in the potential (see figure \ref{fig9} or the right part of figure \ref{twins}). Obviously, the field will roll just for a short distance $\Delta\phi$, which depends on the initial speed $\dot\phi(0)$ that we take from the beginning to be $\dot \phi_{\mathrm{min}}$. Let us first study the simple linear potential
\be\label{straight}
V=\Lambda+c\phi\,,
\ee
where $c$ is a positive slope. Under the simplifying assumption, $\Lambda\gg c\phi$, the solution to the equation of motion is
\be
\Delta\phi(t) \equiv \phi(t)-\phi(0) = -\frac{cM_{Pl}}{\sqrt{3\Lambda}}t + \left(\frac{\dot{\phi}_{\mathrm{min}}M_{Pl}}{\sqrt{3\Lambda}} + \frac{cM^2_{Pl}}{3\Lambda}\right) \left(1-e^{-\sqrt{3\Lambda}t/M_{Pl}}\right)\,.\nonumber
\ee
The term linear in $t$ describes a rolling down at constant speed that eventually dominates the exponentially decreasing term. If the field starts with a positive $\dot{\phi}_{\mathrm{min}} > 0$, it will climb up the hill for a distance
\be \label{delta phi}
\Delta\phi\simeq\frac{\dot\phi_{\mathrm{min}} M_{Pl}}{\sqrt{3\Lambda}}-\frac{cM^2_{Pl}}{3\Lambda}\log\left( 1 + \frac{\sqrt{3\Lambda} \dot\phi_{\mathrm{min}}}{cM_{Pl}} \right)\,,
\ee
in a time
\be \label{delta t}
\Delta t\simeq \frac{M_{Pl}}{\sqrt{3\Lambda}}\log\left(1+\frac{\sqrt{3\Lambda} \dot\phi_{\mathrm{min}}} {cM_{Pl}}\right)
\ee
before it stops and starts rolling down again. The number of e-foldings therefore is generically short unless the slope is exponentially small (if instead $\dot\phi_{\mathrm{min}}$ is very large, then $H$ is no more well approximated by a constant).

We would like to emphasize that an uphill motion is never slow roll. Even if $\epsilon\simeq0=\eta$, when moving uphill $\ddot{\phi}$ is always very large (the motion is hampered both by the slope and by the Hubble friction) and can not be neglected. In fact, the equation of motion is genuinely of second order and the uphill phase depends critically on the initial condition $\dot\phi_0$. On the other hand, the downhill slow-roll motion is an attractor and the solution eventually reaches it independently of $\dot\phi_{\mathrm{min}}$ (of course only if the potential is of the slow-roll type).

We now have all the ingredients to address the question whether the inflaton will overshoot the maximum of the inflaton potential. To this end we describe the part of the potential before the minimum, $\phi>\phi_{\mathrm{min}}$, with the damped oscillator of section \ref{damped} (see figure \ref{fig9}). The key result is eq.~(\ref{phidot in}), the speed of the inflaton $\dot\phi_{\mathrm{min}}$ when it reaches $\phi_{\mathrm{min}}$. We then approximate the uphill phase between maximum and minimum, $\phi_{\rm max} < \phi < \phi_{\rm min}$, by a linear potential. The estimate may seem very rough, but if we take the steepness of our linear potential ($c$ in eq.~(\ref{straight})) to be the maximum steepness reached by the potential $V(\phi)$ between $\phi_{\mathrm{max}}$ and $\phi_{\mathrm{min}}$ then we have an upper bound. If overshooting is possible in this extremal case, then it is also possible for the exact potential.

We use the result eq.~(\ref{phidot in}) as initial condition in eq.~(\ref{delta phi}). One can see that the time $\tm$ in eq.~(\ref{phidot in}) is always smaller than $2M_{Pl}/\sqrt{3\Lambda}$ so that, to get an order of magnitude estimate,  we can neglect the exponential in that formula. Neglecting also numerical factors we take
\be
\dot\phi_{\mathrm{min}}\sim\phi_0\eta\sqrt{3\Lambda}/M_{Pl}\,,
\ee
where $\phi_0 = \phi_i-\phi_{\rm min}$ is the distance between the initial position and the minimum in the damped oscillatory phase of section \ref{damped} (see figure \ref{fig9}). Substituting it into eq.~(\ref{delta phi}), we obtain
\be
\Delta \phi\sim \frac{\dot\phi_{\mathrm{min}}M_{Pl}}{\sqrt{3\Lambda}}\sim\phi_0\eta\,,
\ee
where we have used that $cM^2_{Pl}/3\Lambda\ll \phi_0$, which is generic for our potential. We conclude that overshooting can happen, with $\eta\gtrsim 1$ and a comfortably natural choice $\phi_0\gtrsim\Delta \phi$ (also an initial $\dot\phi\neq 0$ at the beginning of the underdamped oscillatory phase will help to overshoot\footnote{The issue of overshooting for an inflection point potential with particular attention to the role of initial conditions has been recently addressed in \cite{Underwood:2008dh}, where the whole DBI action is taken into account.}). Typically, one does not obtain a large number of e-foldings, see eq.~(\ref{delta t}). We have to remember though that eq.~(\ref{delta t}) is valid just for a linear potential, in our case instead there is maximum, where the slope vanishes.

As an aside we comment on the intriguing correlation between a small cosmological constant and the underdamped oscillatory regime. A graceful exit from inflation typically requires that the inflaton reaches a minimum and starts oscillating and decaying (brane inflation is an interesting exception). In section \ref{damped} we have seen that the underdamped regime, leading to oscillation around the minimum, requires $\eta\gtrsim1$. Equivalently, it requires that the cosmological constant, $\Lambda$, is smaller than the inflaton mass squared $m^2$, in Planckian units. Consider now an inflaton protected by some symmetry that therefore acquires an extremely small mass, e.g.~only from non-perturbative effects. Then an anthropic selection principle would apply: all universes with $\Lambda\gtrsim m^2M^2_{Pl}$ would not have a graceful exit from inflation and would hence be empty.

Summarizing, if the inflaton potential is just $\frac{1}{2}m^2(\phi-\phi_{\rm min})^2$ plus a cosmological constant, then anthropic arguments lead to an upper bound on the cosmological constant of order $\Lambda\lesssim m^2M^2_{Pl}$. An extremely small inflaton mass might then explain the presence of a comparably small cosmological constant which may be responsible for today's measured accelerated cosmic expansion.

It would be interesting to study further features of the inflaton potential eq.~(\ref{effective}). A preliminary observation is that, if an uphill phase is present, a largely non scale-invariant spectrum is produced. The spectral index during the uphill motion is given by
\be
n_s-1\equiv\frac{d\ln \mathrm{P}_{R}}{d\ln k}\simeq -\frac{\dot\phi}{H^2}
-\frac{1}{H}\partial_t\Big(\log\frac{\dot\phi}{H^2}\Big)\,,
\ee
where the quantities on the right-hand side have to be calculated at the time of horizon crossing. After some massage and using the Friedmann equations we obtain
\be
n_s-1\simeq4+\frac{c}{H\dot\phi}-\frac{\dot\phi}{H^2}+\frac{\dot H}{H^2}\,.
\ee
The various terms do not cancel as it happens in the slow-roll regime; the reason can be traced back to the fact that $\ddot\phi$ is not small in this case. Per se, the absence of scale-invariance is not a problem if the perturbations produced during the uphill phase are not those responsible for the CMB inhomogeneities, e.g.~if the uphill phase takes place before or after 60 e-foldings prior to the end of inflation. These issues certainly deserve further study.


\section{Forces on D3- and Anti D3-branes.}\label{sec:anti D3}

In this section we enumerate the contributions to the potential for an anti D3- and a D3-brane (a sketch is given in table~\ref{Contributions} below) and comment on their relative importance.

To summarize: the anti D3-brane is led to the tip ($r\simeq0$) by the interaction with the background; there its angular position is determined by the bulk and moduli stabilization effects. The motion of the D3-brane is governed by moduli stabilization effects (breaking of the no-scale structure). Finally, the attractive Coulomb potential is generically strongly suppressed and plays a role only in very fine tuned or symmetric circumstances.

\paragraph{Background effects.}
We consider the action
\be\label{back}
S_{D3/\overline{D3}}=-T_{D3}\int d^4x\sqrt{-g}\Phi_{\pm},
\ee
where $\Phi$ is defined in terms of the warp factor and the 5-form field strength of $C_4$ as
\be
\Phi_{\pm}\equiv e^{4A}\pm\alpha.
\ee
Therefore in the GKP setup \cite{Giddings:2001yu}, the D3-brane doesn't feel any force (it is BPS with respect to the background). On the contrary, an anti D3-brane tends to fall to the bottom of the (deformed) conifold (small warp factor) to minimize $S_{\overline{D3}}$. As eq.~(\ref{back}) has no angular dependence, the anti D3-brane at the tip enjoys a translational $S^3$ symmetry. The leading contribution to the potential is the warped anti D3-brane tension which can be used \cite{Kachru:2003aw} to break SUSY and uplift an AdS vacuum to a dS vacuum.

\paragraph{Bulk effects.}
To have a compact manifold at a certain radius the conifold has to be cut and glued to a compact Calabi-Yau manifold. Then other ``bulk'' effects for the anti D3-brane arise. These break all the residual symmetry of the conifold as a Calabi-Yau manifold has no continuous symmetry. In \cite{Aharony:2005ez} the warp factor dependence of bulk  effects has been calculated via the AdS/CFT correspondence. The result is that a mass for the anti D3-brane is induced of order
\be \label{bulk}
m_{bulk}^2 \sim (g_sM\al)^{-1}\hh_0^{-0.82} \, ,
\ee
where $M$ is the flux quantum number of the Ramond-Ramond $F_3$-form over the 3-cycle $A$ of the throat and $\hh_0$ is the warp factor at the tip of the throat. Bulk effects would lead the anti D3-brane to a particular angular position in the $S^3$ at the tip. No such effects are present for the D3-brane, again because of its BPS nature with respect to the background.

\paragraph{Moduli stabilization effects.}
To stabilize \label{eta ang}the K\"ahler moduli, one has to break the no-scale structure. Once this is done and the moduli are stabilized (e.g. \`a la KKLT) a mass for the D3-brane open moduli is generated because of their non trivial mixing with the 4-cycle volumes \cite{DeWolfe:2002nn}. The potential generated gives rise to an $\eta$-problem, analogous to the SUGRA $\eta$-problem: it is too step for inflation and generates a ``mass'' for the inflaton of order $H$ \cite{Kachru:2003sx}. This effect is much bigger than the Coulomb attraction and constitutes indeed the leading term of the potential.

As far as the anti D3-brane is concerned, these effects have been investigated in \cite{DeWolfe:2007hd}. They are relevant at the tip of the throat because the background force from eq.~(\ref{back}) does not have an angular dependence. The potential generated by the stabilization of the moduli has the same minima for the D3- and the anti D3-brane \textit{at the tip} (where the anti D3-brane is confined by eq.~(\ref{back})). As the equations for the minimum for D3- and anti D3-branes differ by a term vanishing at $r=0$, away from the tip the respective minima will be generically different.

This effect, together with the bulk one, eq.~(\ref{bulk}), select some vacua in the angular directions at the tip. The relative importance of bulk and stabilization effects depends on the parameters. Comparing the mass from the left-hand side of eq.~(\ref{ang mass}) with eq.~(\ref{bulk}) for the case of the Ouyang (or the simplest Kuperstein) embedding, expanding in $r_0^{3/2}/\mu\ll1$, one gets
\be \label{ineq}
\frac{r_0^{3/2}}{\mu n}\gg \hh_0^{-0.32}\,,
\ee
where $r_0$ indicates the tip of the deformed conifold. As follows from eq.~(\ref{f=1}), the left-hand side of eq.~(\ref{ineq}) has to be much smaller than one to allow to integrate out the stabilized volume and use the remaining effective potential for inflation. Indeed $r_0$ and $\hh_0$ are related by $\hh_0\sim (\sqrt{g_sM}/r_0)^4$, so that the condition to satisfy is
\be
g_sM\gg\frac{(\mu n)^{3/2}}{r_0^{1/3}}\,.
\ee
Although it is possible to fulfil the inequality, this could require a very large flux number $M$, as $g_s\ll1$ and $\mu n\gg 1$.

\begin{table}
\begin{indented}
\lineup
\item[]\begin{tabular}{@{}lll}
\br
Source &$\overline{D3}$&$D3$\cr
\hline\hline
Bulk background   &  $m^2_{bulk} \sim (g_sM\al)^{-1}a_0^{3.29}$ & no effect (BPS) \cr
Throat background & $V\sim 2T_{\overline{D3}}h(w_{\overline{D3}})$ & no effect (BPS) \cr
                  & \textrm{leads $\overline{D3}$ quickly to the tip}
                & \cr
$\rm Coulomb $& $  +V_{D\overline D}$&$-V_{D\overline D}$ \cr
$\rm Tachyon $& develops at $r^2 = {\cal O} (\al)   $ & develops at $r^2 = {\cal O} (\al) $\cr
Moduli stabilization & only known at the tip & $V_{\rm KKLT} + V_{\rm up} + \Delta V$\cr
\br
\end{tabular}
\end{indented}
\caption{\label{Contributions} Contributions to the potential for a D3- and an anti D3-brane.}
\end{table}


\vspace{2mm}

\paragraph{Coulomb potential.}
The Coulomb potential written in terms of the canonically normalized field $\phi$ is
\be \label{coulomb}
V_{\rm up}+V_{D3\overline D3}\simeq-\frac{4\pi^2\phi_0^4}{N}\left(1-\frac{1}{N}\frac{\phi_0^4}{\phi^4} \right).
\ee
where $N=K\,M$ is the product of the fluxes on the 3-cycles of the conifold and $\phi_0$ is the canonically normalized radial position of the anti D3-brane at the tip of the throat. This potential can be obtained considering the D3-brane backreaction on the metric in eq.~(\ref{back}) and keeping the leading order.

We want to argue that the Coulomb interaction is typically weak and subleading (as was noticed for the first time in \cite{Kachru:2003sx}). For example, suppose we use the constant term in eq.~(\ref{coulomb}) for the uplifting $V_{\rm up}$. Then near the minimum in the $\sigma$ direction, $V_{\rm up} \sim V_{\rm AdS}\propto3H^2/(6-\phi^2)^2$ as in eq.~(\ref{Vds}), which estimates the moduli stabilization effects. The Coulomb potential is suppressed with respect to these effects by $(\phi_0/\phi)^4 /N$, as one can see in eq.~(\ref{coulomb}). Therefore as soon as the D3-brane is separated from the anti D3-brane (which sits at the tip $\phi_0$) the Coulomb interaction is by far subleading.


\section{Comments and Conclusions}

We have studied the potential felt by a D3-brane in a warped conifold in the presence of supersymmetrically embedded D7-branes and an anti D3-brane sitting at the tip of the cone. The leading order potential contains three terms: $V_{\rm KKLT}$, an uplifting term $V_{\rm up}$ and $\Delta V$, the latter arising when threshold corrections to the non-perturbative superpotential are taken into account. We have provided general formulae for the extremization in the angular and K\"ahler modulus directions. Once those moduli settle down at their minimum, we are left with an effective potential $V(\phi)$ for the canonically normalized radial D3-brane coordinate $\phi$.

We have studied the possibility to flatten $V(\phi)$ through fine-tuning, such that slow-roll D-brane inflation can be embedded into a type IIB string-theory compactification with all moduli fixed except for the inflaton. We have carried out a detailed analysis for two specific classes of supersymmetric D7-brane embeddings. In the throat (for small $\phi$), $\Delta V$ has a linear term in $\phi$ and otherwise depends on $\phi$ only via integer powers of $\phi^{3/2}$, whereas $V_{\rm KKLT}$ and $V_{\rm up}$ contain terms proportional to $\phi^2$. This means that the potential can be made flat only for a small range of $\phi$. Allowing for fine tuning, a flat inflection point can be generated. In this case the D3-brane dynamics sustains a prolonged stage of slow-roll inflation.

As we do not exactly know how much fine tuning in the effective parameters can be achieved by varying the discrete string theory parameters, we have also considered the issue of whether, for a generic (non fine tuned) shape of the potential, the D3-brane can fall all the way down into the throat where it would annihilate with the anti D3-brane.


The Coulomb attraction which was supposed to drive inflation is generically overwhelmed by the stronger forces generated by the F-term potential. Even in the most promising case, the Kuperstein embedding, where the potential exhibits a flat region around an inflection point, it seems that one will need to take into account, \textit{at the same time}, the Coulomb and the F-term potential. This means that the analyzes based on the simple brane-anti brane potential have to be reviewed if the brane system is embedded into a bona fide string compactification.

A comment on possible further corrections is in order. Quantum corrections, from loop or $\al$ effects, are generically subleading in the KKLT stabilization scenario, the reason being a very small $W_0$ (see e.g.~\cite{Conlon:2005ki}). But the force exerted by the effective potential $V(\phi)$ on the inflaton is hierarchically weaker than the one responsible for the stabilization of the closed string moduli (that is why we can talk about an effective $V(\phi)$ in the first place). Therefore, we expect that that quantum corrections will have sizable effects on warped brane inflation scenarios analyzed here. It would be interesting therefore to study these effects e.g.~along the lines of \cite{vonGersdorff:2005bf,Berg:2005yu,Berg:2007wt}.

In view of the difficulties that the KKLMMT scenario currently presents for the embedding of brane inflation into string theory, it might be interesting to look for qualitatively distinct scenarios such as multi brane inflation \cite{Becker:2005sg}, \cite{Ashoorioon:2006wc} in heterotic M-theory which is based on phenomenologically very interesting flux compactifications \cite{Witten:1996mz}, \cite{Curio:2000dw}, \cite{Curio:2003ur} or related constructions in heterotic string theory \cite{Olsson:2007he}. Other alternatives comprise modular inflation, for instance the racetrack inflation models \cite{Blanco-Pillado:2004ns}, \cite{Lalak:2005hr}, \cite{Blanco-Pillado:2006he} or K\"ahler moduli inflation models \cite{Conlon:2005jm}, \cite{Bond:2006nc}, which have received some attention recently. Furthermore \cite{Thomas:2007sj}, \cite{Spalinski:2007dv}, \cite{Singh:2006yy}, \cite{Cline:2005ty} present interesting ideas towards potentially viable string inflation scenarios.


\ack
It is a pleasure to thank J.~Cline, S.~Kachru, S.~K\"ors, D.~Lyth, P.~Ouyang, A.V.~Ramallo, D.~Tsimpis, A.~Vikman and M. Zagermann for helpful discussions or correspondence. E.~P.~is particularly grateful to M.~Haack for many interesting discussions. This work is supported in part by the European Community's Human Potential Program under contract MRTN-CT-2004-005104 'Constituents, fundamental forces and symmetries of the universe'. A.\ K.\ is supported by the German Research Foundation (DFG) and the Transregional Collaborative Research Centre TRR~33 ``The Dark Universe''. E.\ P.\ is supported by the DFG within the Emmy Noether-Program (grant number: HA 3448/3-1).

\appendix

\section{The Warped Conifold Setup}\label{app:conifold}

The singular conifold is a non-compact Calabi-Yau threefold. It can be defined as a hypersurface in $\mathbb{C}^4$. Two sets of complex projective coordinates are particularly useful: one set is denoted by $w_A$, with $A=1,2,3,4$, and it is the one used to write e.g.~the Ouyang embedding in eq.(\ref{e:O}). In terms of the $w_{A}$, the singular conifold is defined by the equation
\be\label{project}
w_1w_2-w_3w_4=0\,.
\ee
A second set is denoted by $z_A$, used e.g.~to write the Kuperstein embedding in eq.~(\ref{e:K}). The defining equation written in terms of $z_A$ is
\be\label{coni3}
\sum_{A=1}^4(z_A)^2=0\,.
\ee
The two sets of coordinates are linearly related by
\be \label{z}
z_1=\frac12(w_1+w_2)\,, \qquad z_2=\frac{1}{2{\rm i}}(w_1-w_2)\,,
\nonumber \\
z_3=\frac12(w_3-w_4)\,, \qquad z_4=\frac1{2{\rm i}}(w_3+w_4)\,,
\ee
We can also use six real coordinates to parameterize the conifold. The metric is then given by
\be
d s_6^2 &=& d r^2 + r^2 ds_{T^{1,1}}^2 \,,\\
d s_{T^{1,1}}^2 &= &\frac{1}{9} \Bigl(
d \psi + \sum_{i=1}^2 \cos \theta_i \, d \phi_i \Bigr)^2 +
\frac{1}{6} \sum_{i=1}^2 \Bigl( d \theta_i^2 + \sin^2 \theta_i\,d \phi_i^2 \Bigr) \, .
\ee
This makes explicit that the singular conifold has a radial direction $r$ and a base parameterized by 5 angular directions $\phi_1,\,\phi_2,\,\theta_1,\,\theta_2$ and $\psi$. The base is $T^{1,1}$, i.e.~the coset space $(SU(2)_A \times SU(2)_B)/U(1)_{R}$ and it is topologically equivalent to $S^3\times S^2$. The complex $w_A$ coordinates can be expressed in terms of the real coordinates as
\begin{eqnarray}\label{w}
w_1 &=& r^{3/2} e^{\frac{i}{2}(\psi-\phi_1-\phi_2)} \sin
\frac{\theta_1}{2} \sin \frac{\theta_2}{2}\, ,\\
w_2 &=& r^{3/2} e^{\frac{i}{2}(\psi+\phi_1+\phi_2)} \cos
\frac{\theta_1}{2} \cos \frac{\theta_2}{2}\, ,\\
w_3 &=& r^{3/2} e^{\frac{i}{2}(\psi+\phi_1-\phi_2)} \cos
\frac{\theta_1}{2} \sin \frac{\theta_2}{2}\, ,\\
w_4 &=& r^{3/2} e^{\frac{i}{2}(\psi-\phi_1+\phi_2)} \sin
\frac{\theta_1}{2} \cos \frac{\theta_2}{2}\, .
\end{eqnarray}
As for all Calabi-Yau manifolds, the metric of the singular conifold is given by the second derivative of a K\"ahler potential. This is given in terms of the $w_A$ coordinates by
\be\label{Kcon}
     k(w_i,\bar w_i) = r^2 = \left(\sum_{i=1}^4
|w_i|^2\right)^{2/3}\, .
\ee
To obtain the metric one of the four coordinates $\{w_1,w_2,w_3,w_4\}$ has to be expressed in terms of the others using the defining equation eq.~(\ref{project}). If, e.g.~we choose to keep $\{w_2,w_3,w_4\}$ and express $w_1$ as a function of them, the inverse metric (appearing in eq.~(\ref{deltaV})) is given by
\be
k^{\bar \jmath,i}&\equiv& (k_{i,\bar \jmath})^{-1} \\
\nonumber &=& \frac{3}{2} r\Bigg\lbrace
 \delta_{ij}+\frac{w_i\bar w_j}{2r^3}-\frac{1}{r^3}
\left( \begin{array}{ccc}
    |w_1|^2 && \\ &|w_4|^2& \\ &&|w_3|^2
        \end{array}\right)  \\
        &&\hspace{2.5cm} \nonumber
  +\frac{|w_1|^2}{2r^3}\left[\delta_{1i}(\delta_{1j}-1)\frac{\bar
w_i}{\bar
        w_j}+\delta_{1j}(\delta_{1i}-1)\frac{w_i}{ w_j}
\right] \Bigg \rbrace \\[0.5cm]
\nonumber &=& \frac3{2r^2} \left(
\begin{array}{ccc}
r^3-|w_1|^2+\frac12 |w_2|^2 &
\frac{w_3}{w_2}(|w_2|^2+2|w_4|)^2  &
\frac{w_4}{w_2}(|w_2|^2+2|w_3|)^2 \\
    \frac{\bar w_3}{\bar
w_2}(|w_2|^2+2|w_4|)^2  &r^3-|w_4|^2+\frac12 |w_3|^2 &  w_4\bar
w_3 \\
    \frac{\bar w_4}{\bar w_2}(|w_2|^2+2|w_3|)^2 &   w_3 \bar
w_4         & r^3-|w_3|^2+\frac12 |w_4|^2
\end{array}
\right)\,,
\ee
where $i$ and $j$ run from 1 to 3.



\section{On the Parameters}\label{a:p}

The parameters of the models are the following
\begin{itemize}
\item $\mu$: represents the deepest $r_{D7}$ value reached by the D7-brane. We require that $\mu \gg r^{3/2}$ so that the stabilized volume does not change much during the D3-brane radial motion.
\item $A_0$: the complex structure dependent factor in $W_{np}$. Its phase can be absorbed by a shift of the axion. Once $W_0$ is chosen as in eq.~(\ref{param2}) and $D$ as in eq.~(\ref{paramb}), $A_0$ can be factorized out of the scalar potential. Therefore it does not play an important role in the discussion.
\item $W_0$: its value can be fine tuned (up to a certain precision) varying the fluxes which fix the complex structure moduli and the dilaton. The constraint on its value comes from the KKLT procedure for fixing the overall volume (generically all K\"ahler moduli). This generically requires $W_0 \ll 1$.
\item $\beta$: it fixes the value of the vacuum energy. When a particular uplifting procedure is specified, its value is determined in terms of stringy parameters (for example the position of the anti D3-brane at the tip of the throat or the world volume fluxes in a D-term uplifting). Having a de Sitter vacuum translates into $\beta\gtrsim1$, having a minimum at all requires $\beta\lesssim\mathcal{O}(4)$. This parameter is important for the shape of the effective potential $V(\phi)$. Only a special fine tuned value of $\beta$ leads to a flat inflection point.
\end{itemize}


\section{Dependence of $\scr$ on Uplifting and Inflaton}\label{scr dep}

The scalar potential
\be
V_{\rm AdS}=\frac{aA_0e^{-a\sigma}}{2\sigma^2}\left(\frac{1}{3}\sigma a A_0 e^{-a\sigma}+W_0+A_0e^{-a\sigma} \right),
\ee
resulting from
\be
K&=&-3\mathrm{log}(\rho+\overline \rho)\,,\\
W&=&W_0+A_0\,e^{-a\rho}\,,
\ee
has the well known \cite{Kachru:2003aw} AdS minimum $\sigma_0$, given by the solution of the transcendental equation
\be \label{min ads}
\sigma_0=-\frac{3}{2} \frac{A_0+W_0\,e^{a\sigma_0}}{aA_0},
\ee
where $W_0$ is a negative real number. The aim of this appendix is to calculate how the minimum $\sigma_0$ changes when the uplifting and the D3-brane dynamics are taken into account.


\subsection {Shift of Critical Volume through Uplifting} \label{scr up}

For concreteness we look at an anti D3-brane uplifting. If an anti D3-brane is present in the ISD solution of \cite{Giddings:2001yu}, it will feel a potential that pulls it to the tip of the throat. Supersymmetry is broken and the effective scalar potential receives a contribution proportional to the redshifted anti D3-brane tension that we schematically write as
\be
V_{\rm up}=\frac{D}{4\sigma^2}.
\ee
The potential's minimum will consequently shift from $\sigma_0$ to $\sd=\sigma_0+\Delta_\beta$, where $\Delta_{\beta} \ne 0$. It is useful to trade the parameter $D$ for another parameter $\beta$, which is introduced via
\be \label{def D}
D = \frac23 \beta \sigma_0 a^2A_0^2e^{-2a\sigma_0}\,.
\ee
The condition that $V_{\rm up}$ uplifts the AdS minimum to dS is now easily expressed through the requirement that $\beta\gtrsim 1+2\Delta_\beta/\sigma_0$ (which is very close to, but not exactly one). In what follows we assume that this condition is fulfilled and therefore the minimum is dS. The equation $\partial_{\sigma}(V_{\rm AdS}+V_{\rm up}) = 0$ leads to the critical point
\be \label{trasc up}
\sd=-\frac{1}{4aA_0}\left[ 7A_0+3W_0e^{a\sd}-\sqrt{(A_0-3W_0e^{a\sd})^2-\frac{24De^{a\sd}}{a}} \right]\,.
\ee
This is, like eq.~(\ref{min ads}), a transcendental equation and has to be solved numerically. To get some analytical control, we use the following trick. We substitute $D$ and $W_0$ in eq.~(\ref{trasc up}) using eq.~(\ref{def D}) and eq.~(\ref{min ads}). Then we solve the resulting equation for $\Delta_\beta$. This can be done by expanding $e^{a\Delta_\beta} \simeq 1+a\Delta_\beta$ so that the equation is no longer transcendental. Eventually, the only transcendental equation that we have to solve numerically is eq.~(\ref{min ads}), and we arrive at an analytical expression for $\Delta_\beta$.

The resulting expression for $\Delta_\beta$ is a bit long, so we write its expansion in $1/a\sigma_0$, which reads
\be
\Delta_{\beta}\simeq\frac{\beta}{a^2\sigma_0}+\frac{\beta(4\beta-5)}{2a^3\sigma_0^2}+\dots\,,
\ee
in very good agreement with the numerical calculation. We note that this is equivalent to an expansion in $D/(aW_0^2)$ of eq.~(\ref{trasc up}); in fact from eq.~(\ref{def D}) one sees that $D$ is suppressed with respect to $W_0^2$ by a factor $1/\sigma_0$. This expansion would give the transcendental equation
\be \label{min up}
\sd=-\frac32 \frac{A_0+W_0e^{a\sd}}{aA_0}-\frac{3De^{2a\sd}}{a^2A_0(A_0-3W_0e^{a\sd})}+\dots\,.
\ee


\subsection{Shift of Critical Volume through D3-brane}

In this section we take into account also a dynamical D3-brane and calculate its effect on the minimum of the potential in the $\sigma$ direction that we call $\scr$. The potential is given in eq.~(\ref{pot tot}). We expand $\partial_{\sigma}V=0$ for small $r$ (again this is an $r^2/\sigma$ or an $r^{3/2}/(\mu n)$ expansion). Solving for $\scr$, one finds
\be
\scr(D,r) &=& \scr(r=0)+\scr^{(1)}r+\scr^{(3/2)}r^{3/2}+\dots\nonumber\\
&=&\sd+\frac{9(A_0+3W_0e^{a\scr})}{8a^2\mu^2 n^2 \gamma (A_0-3W_0e^{a\scr})}\,r \nonumber \\
&&-\frac{3(A_0^2+2A^2_0W_0e^{a\scr}+3W_0^2e^{2a\scr})}{2aA_0\mu n (A_0-3W_0e^{a\scr})}\,r^{3/2}+\dots
\ee
where $\scr(r=0)\equiv \sd$ coincides with the critical volume investigated in the previous section and in $\scr^{(1)}$ and $\scr^{(3/2)}$ we have neglected terms suppressed by a factor of order $D/W_0^2$ (see eq.~(\ref{def D})). As we did in the last section we substitute $D$ and $W_0$, using eq.~(\ref{def D}) and eq.~(\ref{min ads}). Then we solve for $\Delta_{r}=\scr-\sd$. The result is
\be
\Delta_{r}=\frac{r^{3/2}}{a\mu n}+\dots
\ee
To summarize, we have estimated analytically the dependence of the minimum on the uplifting and on the D3-brane position; this is at leading order
\be
\scr&=&\sigma_0+\Delta_{\beta}+\Delta_{r}\nonumber \\
&\simeq&\sigma_0+\frac{\beta}{a^2\sigma_0}+\frac{r^{3/2}}{a\mu n}+\dots
\ee


\section{Sign of $r^{3/2}$ Term}\label{r3/2}

In this appendix we show that the expansion of the scalar potential has a negative term at order $r^{3/2}$. This term determines the negative curvature of the potential for small $r$. In fact, the inflaton potential eq.~(\ref{r exp}) also has a term proportional to $r$, which, however, does not contribute to the curvature.

The explicit values for $\Vds^{(3/2)}$ and $\Delta V^{(3/2)}$ are
\be
\Vds^{(3/2)}&=& -\frac{\left[3a A_0^2 e^{-2 a \sigma_0} (a\scr+6)+9D+18A_0aW_0e^{-a\sigma_0}\right]}{18a\mu n\sigma_0^3} \label{v3/2} \, ,  \\
\Delta V^{(3/2)}&=& - \frac{A^2_0ae^{-2a\sigma_0}}{4\mu n\sigma_0^2}\, .
\ee
To see that $V^{(3/2)} = \Vds^{(3/2)}+\Delta V^{(3/2)}<0$ we substitute eq.~(\ref{def D}) and eq.~(\ref{min ads}) into eq.~(\ref{v3/2}) and expand in $\Delta_\beta = \sd-\sigma_0$. This gives
\be
\Vds^{(3/2)}&\simeq&  \frac{A^2_0ae^{-2a\sigma_0}(3-2\beta)}{6\mu n\sigma_0^2}+\dots
\ee
Therefore
\be
\frac{\Vds^{(3/2)}}{\Delta V^{(3/2)}}\simeq-\frac{4(3-2\beta)}{6}\gtrsim -1/2 \,,
\ee
for $\beta\gtrsim 1$. We are thus left with
\be
V^{(3/2)}&=&-\frac{A_0^2ae^{-2a\sigma_0}}{12\sigma_0^2\mu n}(4\beta-3)<0\, ,
\ee
in agreement with eq.~(\ref{C3/2}).


\section{Maximum and Minimum of $V(\phi)$} \label{max min}

\begin{figure}
\centering
\includegraphics[height =0.4 \textwidth, width=0.7 \textwidth]{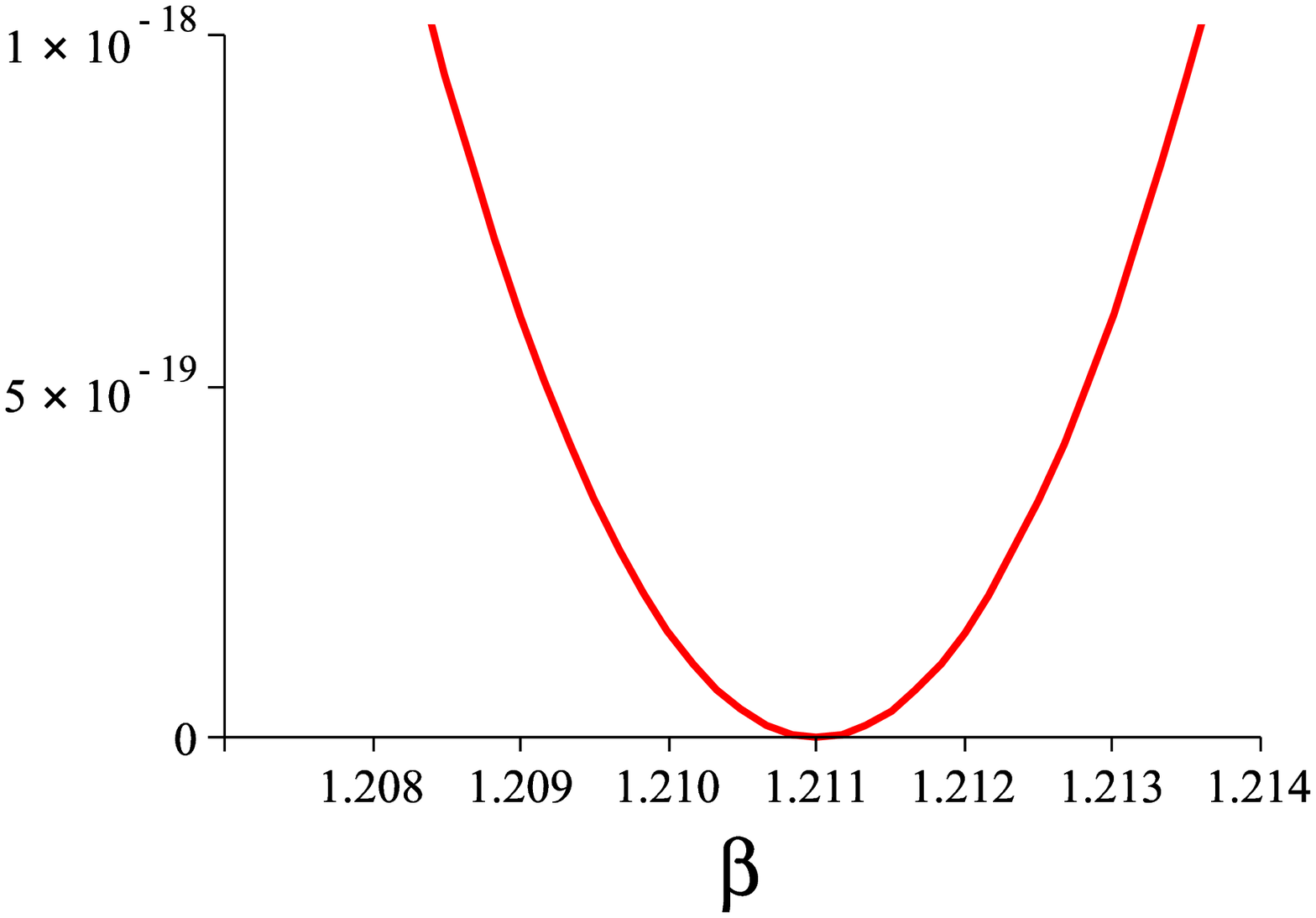}
\caption{\label{infle}The plot shows the discriminant,
eq.~(\ref{discri}), as a function of $\beta$ (without performing the $\Delta/\sigma_0$ expansion). When the discriminant is zero the minimum and maximum of the potential coincide and we get a flat inflection point.}
\end{figure}

In this section we show that in the case of the Kuperstein embedding, the effective potential in the $r$ (canonically $\phi$) direction has always a maximum and a minimum (in a special case they coincide). Our starting point is the obtained inflaton potential
\be
V(\phi)&=&\frac{A_0^2a^2e^{-2a\sigma_0}}{6\sigma_0}(\beta-1) + \phi\frac{9A_0^2e^{-2a\sigma_0} M^2_{Pl}}  {16T^{3/2}_{D3}\sigma_0^{3/2}\mu^2n^2}\nonumber \\
&&-\phi^{3/2}\frac{A_0^2ae^{-2a\sigma_0}}{12\sigma_0^{5/4}T_{D3}^{3/4}\mu n}(4\beta-3)+\frac{\phi^2}{3M^2_{Pl}}\frac{A_0^2a^2e^{-2a\sigma_0}}{6\sigma_0}(\beta-1)\,,
\ee
which is the potential eq.~(\ref{effective}) with
\be
\Lambda&=&\frac{A_0^2a^2e^{-2a\sigma_0}}{6\sigma_0}(\beta-1)\,,\nonumber \\
C_1&=&\frac{9A_0^2e^{-2a\sigma_0} M^2_{Pl}}  {16T^{3/2}_{D3}\sigma_0^{3/2}\mu^2n^2}\,, \nonumber\\
C_{3/2}&=&\frac{A_0^2ae^{-2a\sigma_0}}{12\sigma_0^{5/4}T_{D3}^{3/4}\mu n}(4\beta-3)\,,\label{C3/2} \\
C_2&=&\frac{M_{Pl}^2}{3}\Lambda \nonumber \; .
\ee
The first derivative of the potential is a quadratic polynomial in $\sqrt{\phi}$. There are two extrema (a maximum and a minimum) when the discriminant is positive, i.e.
$9C_{3/2}-32C_1C_2>0$. Explicitly, the discriminant reads
\be\label{discri}
9C_{3/2}-32C_1C_2=\frac{A_0^2a^2e^{-2a\sigma_0}}{64\sigma_0^{5/2}T_{D3}^{3/2}\mu^2 n^2}(4\beta-5)^2 \ge 0 \,.
\ee
This quantity is always {\em non-negative} so that $V(\phi)$ will always possess a minimum and a maximum. It is evident that for a particular value of $\beta$ the discriminant becomes zero in which case the maximum and minimum coincide and a flat inflection point occurs. We have plotted the discriminant, without expanding it, in figure \ref{infle}. The plot confirms the result of our leading order calculation.\\


\end{document}